\documentclass[9pt,journal]{IEEEtran}

\pdfoutput=1

\usepackage{cite}

\ifCLASSINFOpdf

\else

\fi

\usepackage[cmex10]{amsmath}

\usepackage{amssymb}

\usepackage{paralist}

\usepackage{pstool}

\usepackage{tikz}
\usetikzlibrary{shapes,arrows,fit}

\newtheorem{thm}{Theorem}
\newtheorem{prop}[thm]{Proposition}

\newtheorem{lem}[thm]{Lemma}
\newtheorem{defn}[thm]{Definition}
\newtheorem{rem}[thm]{Remark}
\newtheorem{example}[thm]{Example}

\providecommand{\norm}[1]{\left\lVert#1\right\rVert}

\DeclareMathOperator{\im}{Im}

\providecommand{\expec}[1]{\mathbf{E}\left[#1\right]}
\providecommand{\prob}[1]{\mathbf{P}\left\{#1\right\}}

\newcommand\tpose{\mathsf{T}}

\usepackage{array}

\begin{document}

\title{Stability and disturbance attenuation for a switched Markov jump linear system}

\author{Collin~C.~Lutz$^\ast$,~\IEEEmembership{Student~Member,~IEEE,} and~Daniel~J.~Stilwell,~\IEEEmembership{Member,~IEEE}\\%
Bradley Department of Electrical \& Computer Engineering\\%
302 Whittemore (0111)\\%
Virginia Tech\\%
Blacksburg, VA 24061\\%
Tel:~(540)~231-3204, Fax:~(540)~231-3362, Email:~\{collin,stilwell\}~at~vt.edu%
\thanks{$^\ast$Corresponding author}%
\thanks{This research was made with Government support under and awarded by DoD, Air Force Office of Scientific Research, National Defense Science and Engineering Graduate (NDSEG) Fellowship, 32 CFR 168a.}%
}

\maketitle

\begin{abstract}
We address a class of Markov jump linear systems that are characterized by the underlying Markov process being time-inhomogeneous with a priori unknown transition probabilities. Necessary and sufficient conditions for uniform stochastic stability and uniform stochastic disturbance attenuation are reported. In both cases, conditions are expressed as a set of finite-dimensional linear matrix inequalities that can be solved efficiently.
\end{abstract}

\IEEEpeerreviewmaketitle

\section{Introduction}

\IEEEPARstart{A}{ discrete-time} Markov jump linear system is a stochastic discrete-time linear time-varying system where the time-variation of system matrices is determined by a realization of a Markov chain. The Markov chain may be time-homogeneous (characterized by constant transition probabilities) or time-inhomogeneous (characterized by time-varying transition probabilities), and we slightly abuse terminology by referring to a time-(in)homogeneous Markov jump linear system. We address a \emph{switched Markov jump linear system}, which is simply a time-inhomogeneous Markov jump linear system where the underlying Markov chain is characterized by an a priori unknown sequence of transition probability matrices that assume one of finitely-many values at each time instant; an a priori unknown \emph{switching sequence} parameterizes the transition probability matrices at each time instant. As a special case, our analysis also applies to time-inhomogeneous Markov jump linear systems with \emph{known} transition probabilities that vary in a finite set. In the existing literature, stability (resp. disturbance attenuation) of a time-inhomogeneous Markov jump linear system is equivalent to an infinite-dimensional Lyapunov (resp. storage function) criterion that in general lacks a practical technique for solving. For stochastically stable and contractive systems, we show the existence of Lyapunov and storage functions with finite dependence on the future. This observation leads to necessary and sufficient conditions for uniform stochastic stability and uniform stochastic disturbance attenuation for a switched Markov jump linear system, expressed as a set of finite-dimensional linear matrix inequalities (LMIs), which can be solved efficiently using well-known techniques.

The Markov jump linear system abstraction finds application in many areas including economics~\cite{lutzStilwell:doVal1999}, fault-tolerant control~\cite{lutzStilwell:chizeck1982diss}, energy-aware control~\cite{lutzStilwell:lutz2013}, and networked control~\cite{lutzStilwell:hespanha2007,lutzStilwell:seiler2005}. Despite the prevalence of Markov jump linear systems, little attention has been paid to the case when the Markov chain transition probabilities are time-varying, and almost no attention has been paid to the case when the Markov chain transition probabilities are time-varying and a priori unknown. 

Time-varying Markov chain transition probabilities may arise in a variety of situations. Consider, for example, a control system where the plant and controller are connected via a wireless communications network subject to random network delays and/or packet loss (see, e.g.,~\cite{lutzStilwell:xiao2000}). Network delays and packet loss probabilities are influenced by many factors, including ambient noise, distance between wireless nodes, the presence of other wireless communication nodes on the same network, and sources of interference on the same frequency band~\cite{lutzStilwell:ploplys2004}. Thus, network delay and packet loss probabilities may vary with time due to, e.g., solar activity, mobile network nodes, evolving network topology, or adversarial disruption (jamming). In some of these scenarios, the time-varying Markov chain transition probabilities may be known in advance, while in other scenarios, the time-variation may be a priori unknown.

Time-homogeneous Markov jump linear systems have been studied quite extensively in the literature. Ji and Chizeck~\cite{lutzStilwell:ji1991} study various second moment stability concepts and the almost sure asymptotic stability for the time-homogeneous case. Costa and Fragoso~\cite{lutzStilwell:costa1993} provide coupled linear matrix inequality conditions equivalent to mean square stability, and Ji and Chizeck~\cite{lutzStilwell:ji1990a} characterize the jump linear quadratic Gaussian optimal control problem. More recently, Seiler and Sengupta~\cite{lutzStilwell:seiler2003} consider the $H_\infty$ control problem and provide a stochastic bounded real lemma that can be used when the Markov chain is time-homogeneous. Lee and Dullerud provide necessary and sufficient conditions for a time-homogeneous Markov jump linear system to be almost surely uniformly exponentially stable~\cite{lutzStilwell:leeDullerud2006Stab} and almost surely uniformly strictly contractive~\cite{lutzStilwell:leeDullerud2006Perf}.

Time-inhomogeneous Markov jump linear systems have received less attention. For the case when the time-varying Markov chain transition probabilities are known, Krtolica et al.\ \cite{lutzStilwell:krtolica1994} provide a necessary and sufficient condition for mean square stability in the form of an infinite set of coupled matrix equations, and Aberkane~\cite{lutzStilwell:aberkane2013} states a similar stability result in terms of an infinite set of LMIs. Fang and Loparo~\cite{lutzStilwell:fang2002} reduce the infinite set of matrix equations in~\cite{lutzStilwell:krtolica1994} to a finite set when the transition probabilities of the Markov chain are periodic. Aberkane~\cite{lutzStilwell:aberkane2013} provides a necessary and sufficient condition for stochastic disturbance attenuation in the form of an infinite set of LMIs, which reduces to a finite set when the transition probabilities of the Markov chain are periodic.

For the case when the Markov chain transition probabilities are time-varying and a priori unknown, only sufficient conditions for uniform stochastic stability and uniform stochastic disturbance attenuation have been provided in the literature. Bolzern et al.\ \cite{lutzStilwell:bolzern2010} examine a continuous-time Markov jump linear system with time-varying a priori unknown Markov process transition rates (see, e.g.,~\cite[Sec. 11.4.2]{lutzStilwell:garcia2008book}) and provide a sufficient condition for uniform stochastic stability subject to a dwell-time constraint. Lutz and Stilwell~\cite{lutzStilwell:lutz2013} examine a particular class of time-inhomogeneous Markov jump linear systems with a priori unknown transition probabilities, provide sufficient conditions for uniform mean square stability and uniform stochastic disturbance attenuation, and present a sufficient condition for uniform stochastic stability subject to an average dwell-time constraint. 

\subsection*{Notation}
The positive and nonnegative integers are represented by $\mathbb{N}$ and $\mathbb{N}_0$, respectively. The standard Euclidean vector norm and corresponding induced matrix norm are both denoted by $\norm{\cdot}$. The set of $n\times n$ symmetric matrices is denoted by $\mathbb{S}_{n}$, and $\mathbb{S}_{n}^{+}$ denotes the set of positive definite symmetric matrices. For an $n\times n$ symmetric matrix $X$, the notation $X>0$ simply means $X\in\mathbb{S}_{n}^{+}$, and $X<0$ means $-X>0$. The set of $N\times N$ \emph{stochastic matrices} is denoted $\mathbb{T}_{N}$ and consists of matrices with nonnegative elements where each row sums to one. The composition of two functions $f,g$ is denoted by $f\circ g$, and the image of a function $f$ is written $\im f$. A \emph{sequence} is a function $f$ whose domain is a subset of the set of integers (usually $\mathbb{N}$ or $\mathbb{N}_0$) and may be equivalently viewed as an ordered list $(f(1),f(2),\dots)$. Given two sets $\mathcal{N}$ and $\mathcal{J}$, the Cartesian product is denoted by $\mathcal{N}\times\mathcal{J}$, and the Cartesian power of a set $\mathcal{J}$ is denoted by $\mathcal{J}^M$ where $M\in\mathbb{N}$. By convention, if $M=0$ then $\mathcal{J}^M=\{\emptyset\}$, a singleton set (e.g., see~\cite[p. 57]{lutzStilwell:hrbacek1999book}), and $\mathcal{N}\times \mathcal{J}^M=\mathcal{N}$. The space of mean square summable stochastic processes on a probability space $(\Omega,\mathfrak{F},\mathbf{P})$ is defined $\ell^2_e = \{ w:\mathbb{N}_0\times\Omega\to\mathbb{R}^m : \norm{w}_{2,e} < \infty \}$ where $\lVert w \rVert_{2,e}^2=\sum_{k=0}^\infty\mathbf{E}[\,\lVert w(k)\rVert^2\,]$.

\section{Preliminaries}\label{sec:prelims}
In this section, we review definitions and results for a Markov jump linear system where the Markov chain is time-inhomogeneous and characterized by known time-varying transition probabilities. Fix a probability space $(\Omega,\mathfrak{F},\mathbf{P})$ and let $\theta:\mathbb{N}_0\times\Omega\to\mathcal{N}$ be a Markov chain defined on the probability space which takes values in a finite set $\mathcal{N}=\{1,\dots,N\}$. For $k\in\mathbb{N}$, define $p_{ij}(k)=\mathbf{P}\{\theta(k)=j\mid\theta(k-1)=i\}$ and let $P(k)$ be the $N\times N$ matrix with entries $p_{ij}(k)$. The Markov chain $\theta$ is \emph{time-homogeneous} if $P$ is constant. Otherwise, the Markov chain is \emph{time-inhomogeneous}. Let $p_i(k) = \mathbf{P}\{\theta(k)=i\}$ and define the row vector $p(k) = [ p_1(k) \; p_2(k) \; \cdots \; p_N(k) ]$. Note that $p(k) = p(0) P(1) P(2)\cdots P(k)$. The initial distribution $p(0)$ and the sequence $P:\mathbb{N}\to\mathbb{T}_{N}$ of transition probability matrices fully specifies the underlying probability structure of $\theta$. Let $\mathcal{G}=\{(A(i),B(i),C(i),D(i)): i \in\mathcal{N}\}$ be a finite set of matrices where $A(i)\in\mathbb{R}^{n\times n}$, $B(i)\in\mathbb{R}^{n\times m}$, $C(i)\in\mathbb{R}^{p\times n}$, and $D(i)\in\mathbb{R}^{p\times m}$. The discrete-time Markov jump linear system, denoted $(\mathcal{G},P,p(0))$, is defined by the difference equation
\begin{IEEEeqnarray}{c}
\begin{bmatrix}
x(k+1) \\
z(k)
\end{bmatrix}
= 
\begin{bmatrix}
A(\theta(k)) & B(\theta(k)) \\
C(\theta(k)) & D(\theta(k))
\end{bmatrix}
\begin{bmatrix}
x(k) \\
w(k)
\end{bmatrix}
 \IEEEeqnarraynumspace\label{eq:mjls}
\end{IEEEeqnarray}
and initial condition $x(0)$, where $x(k)\in\mathbb{R}^n$ is the state vector, $w(k)\in\mathbb{R}^m$ is a disturbance vector, and $z(k) \in\mathbb{R}^p$ is an error vector.

Define the (random) state transition matrix $\Phi(k,j) = A(\theta(k-1))A(\theta(k-2))\cdots A(\theta(j))$ when $k>j$, $\Phi(k,j) = I$ when $k=j$, and $\Phi(k,j)$ is undefined for $k<j$. The zero-input ($w\equiv 0$) solution of the state in~\eqref{eq:mjls} is $x(k) = \Phi(k,0)x(0)$.

\begin{defn}[Def. 3 of~\cite{lutzStilwell:seiler2003}]
The Markov jump linear system $(\mathcal{G},P,p(0))$ is \emph{weakly controllable} if for every initial $(x_0,\theta_0)$ and any final $(x_f,\theta_f)$, there exists a finite time $k_f$ and an input $w_c$ such that $\prob{x(k_f)=x_f, \theta(k_f)=\theta_f \mid x(0)=x_0,\theta(0)=\theta_0}>0$.
\end{defn}

Among the various ways to address stochastic stability for Markov jump linear systems, we find that mean square stability is most appropriate for our approach.
\begin{defn}\label{defn:emss}
The Markov jump linear system $(\mathcal{G},P,p(0))$ is \emph{exponentially mean square stable} if there exist $c\geq 1$ and $0\leq\lambda<1$ such that $\expec{\Phi^{\tpose}(k,j)\Phi(k,j)\mid\theta(j)=i}\leq c\lambda^{k-j}I$ for all $i\in\mathcal{N}$ and for all $k,j\in\mathbb{N}_0$ such that $k\geq j\geq 0$.
\end{defn}

At least two notions of disturbance attenuation for Markov jump linear systems have been examined in the literature~\cite{lutzStilwell:seiler2003,lutzStilwell:leeDullerud2006Perf}. We find that mean square attenuation best suits our approach.
\begin{defn}\label{defn:mssc}
The Markov jump linear system $(\mathcal{G},P,p(0))$ is \emph{mean square strictly contractive} if there exists $\gamma\in(0,1)$ such that whenever $x(0)=0$, $\norm{z}_{2,e} \leq \gamma \norm{w}_{2,e}$ for all $w\in\ell^2_e$.
\end{defn}

The following matrix-valued function will appear extensively in characterizations of disturbance attenuation.
\begin{defn}
 Let $\mathcal{G}$ be given. For $i\in\mathcal{N}$ and $X,Y\in\mathbb{S}_{n}$, define
 \begin{IEEEeqnarray*}{c}
  \mathcal{B}(i,X,Y) = 
\begin{bmatrix}
A(i) & B(i) \\
C(i) & D(i)
\end{bmatrix} ^{\tpose}
\begin{bmatrix}
X & 0 \\
0 & I
\end{bmatrix}
\begin{bmatrix}
A(i) & B(i) \\
C(i) & D(i)
\end{bmatrix}
-
\begin{bmatrix}
Y & 0 \\
0 & I
\end{bmatrix}
 \end{IEEEeqnarray*}
\end{defn}

Exponential mean square stability of the time-inhomogeneous Markov jump linear system $(\mathcal{G},P,p(0))$ with known transition probabilities may be characterized by a stochastic Lyapunov criterion.
\begin{prop}[Thm. 2 of~\cite{lutzStilwell:krtolica1994}, Prop. 1 of~\cite{lutzStilwell:aberkane2013}]\label{prop:lyapTih}
The time-inhomogeneous Markov jump linear system $(\mathcal{G},P,p(0))$ is exponentially mean square stable if and only if there exist $\eta,\rho,\nu>0$ and a function $X:\mathcal{N}\times \mathbb{N}_0\to \mathbb{S}_{n}^{+}$ such that $\eta I \leq X(i,k) \leq \rho I$ and $A^{\tpose}(i) \tilde{X}(i,k+1) A(i) - X(i,k) \leq  -\nu I$ for all $i\in\mathcal{N}$ and all $k\in\mathbb{N}_0$ where $\tilde{X}(i,k+1)=\sum_{j=1}^N p_{ij}(k+1)X(j,k+1)$. Moreover, if $X$ exists, one may take $c=\rho/\eta$ and $\lambda=1-\nu/\rho$ in Definition~\ref{defn:emss}.
\end{prop}

Proposition~\ref{prop:lyapTih} is a stochastic version of the familiar Lyapunov stability criterion (e.g.,~\cite[Thm. 23.3]{lutzStilwell:rugh1996book}) for discrete-time linear time-varying systems. If $X$ exists in Proposition~\ref{prop:lyapTih}, then $V(i,k,y) := y^{\tpose}X(i,k) y$ is a uniformly positive and uniformly bounded stochastic Lyapunov function for $(\mathcal{G},P,p(0))$ and satisfies $\mathbf{E} [ V(\theta(k+1),k+1,x(k+1)) - V(\theta(k),k,x(k)) \mid x(k)=y, \theta(k)=i ] \leq  -\nu \lVert y \rVert^2$ for all $i\in\mathcal{N}$, $k\in\mathbb{N}_0$, and $y\in\mathbb{R}^n$. Of course, the utility of Proposition~\ref{prop:lyapTih} for assessing stability of a given system is limited due to the infinite number of matrices being prohibitively difficult to compute in practice.

Disturbance attenuation for a time-inhomogeneous Markov jump linear system can also characterized by an infinite set of LMIs.
\begin{prop}[Thm. 1 of~\cite{lutzStilwell:aberkane2013}]\label{prop:tihBrl}
Assume $(\mathcal{G},P,p(0))$ is weakly controllable and $p_i(k)>0$ for all $i\in\mathcal{N}$ and $k\in\mathbb{N}_0$. The time-inhomogeneous Markov jump linear system $(\mathcal{G},P,p(0))$ is exponentially mean square stable and mean square strictly contractive if and only if there exist $\eta,\rho,\nu>0$ and a function $X:\mathcal{N}\times \mathbb{N}_0\to \mathbb{S}_{n}^{+}$ such that $\eta I  \leq  X(i,k)  \leq  \rho I$ and $\mathcal{B}(i,\tilde{X}(i,k+1),X(i,k)) \leq  -\nu I$ for all $i\in\mathcal{N}$ and all $k\in\mathbb{N}_0$ where $\tilde{X}(i,k+1) = \sum_{j=1}^N p_{ij}(k+1) X(j,k+1)$.
\end{prop}

Proposition~\ref{prop:tihBrl} is a stochastic version of the Kalman-Yakubovich-Popov (KYP) lemma (e.g., see~\cite[Cor. 12]{lutzStilwell:dullerud1999}) for discrete-time linear time-varying systems. If $X$ exists in Proposition~\ref{prop:tihBrl}, then $V(i,k,y) := y^{\tpose} X(i,k) y$ is a uniformly positive and uniformly bounded \emph{stochastic storage function} (see~\cite{lutzStilwell:willems1972i}) for $(\mathcal{G},P,p(0))$ and satisfies $\mathbf{E} [ V(\theta(k+1),k+1,x(k+1)) - V(\theta(k),k,x(k)) + \lVert z(k) \rVert^2 ] \leq \gamma^2 \mathbf{E} [ \lVert w(k) \rVert^2 ]$ for all $i\in\mathcal{N}$, $k\in\mathbb{N}_0$, $y\in\mathbb{R}^n$, and $w\in\ell^2_e$. Again, the utility of Proposition~\ref{prop:tihBrl} for assessing disturbance attenuation properties of a given system is limited due to the infinite number of matrices being prohibitively difficult to compute in practice.

\section{Switched Markov jump linear system}\label{sec:smjls}
In this section, we examine a time-inhomogeneous Markov jump linear system with a priori unknown time-varying transition probabilities. We assume that the sequence $P$ of transition probability matrices is not known in advance but takes values in some finite set of matrices $\{\Pi(1),\dots,\Pi(J)\}$ where $\Pi(s)\in\mathbb{T}_{N}$ for $s\in\mathcal{J}=\{1,\dots,J\}$. Thus, $P(k)=\Pi(\psi(k))$ for some a priori unknown sequence $\psi:\mathbb{N}\to\mathcal{J}$. The notation $\pi_{ij}(\psi(k))$ denotes the $ij$-th element of matrix $\Pi(\psi(k))$.

A \emph{switched Markov jump linear system}, denoted $(\mathcal{G},\Pi,\Psi,p(0))$, is defined to be the collection of Markov jump linear systems $\{ (\mathcal{G} , \Pi \circ \psi, p(0)) : \psi\in \Psi \}$ where $\Psi$ is the application-specific set of all sequences that may occur. Depending on the application, $\Psi$ could be the set of \emph{all} sequences in $\mathcal{J}$. Alternatively, some applications may disallow certain sequences from occurring due to problem-specific information available. Each member of the switched Markov jump linear system is driven by a different time-inhomogeneous Markov chain with transition probabilities given by $\Pi(\psi(k))$, $k\in\mathbb{N}$ for some sequence $\psi\in\Psi$. The \emph{switched} modifier here is used to draw analogy to deterministic switched systems (see, e.g.,~\cite{lutzStilwell:liberzon2003book}), and we often refer to $\psi$ as a switching sequence. For $M\in\mathbb{N}$ and $k\in\mathbb{N}_0$, we define $\psi_{M}(k) = (\psi(k+1),\psi(k+2),\dots,\psi(k+M))$. Additionally, the set of all sequences of length $M$ that occur in $\Psi$ is denoted $\Psi_{M} = \left\{ \psi_{M}(t) : \psi\in\Psi, t\in\mathbb{N}_0 \right\}$ and is a subset of $\mathcal{J}^M$.

\subsection{Stability}
Since we now address time-inhomogeneous Markov jump linear systems where the sequence of transition probability matrices is not known a priori, we modify the definition of stability so that it applies uniformly over all possible sequences of transition probability matrices.
\begin{defn}\label{defn:uemss}
The switched Markov jump linear system $(\mathcal{G},\Pi,\Psi,p(0))$ is \emph{uniformly exponentially mean square stable} if there exist $c\geq 1$ and $0\leq\lambda<1$ such that $\mathbf{E} [ \Phi^{\tpose}(k,j) \Phi(k,j)\mid\theta(j)=i ] \leq c\lambda^{k-j}I$ for all $i\in\mathcal{N}$, all $k,j\in\mathbb{N}_0$ such that $k\geq j\geq 0$, and all $\psi\in\Psi$. 
\end{defn}

Uniformity in Definition~\ref{defn:uemss} refers to the uniform decay rate for all $\psi\in\Psi$. Thus, uniform exponential mean square stability ensures that each individual Markov jump linear system in the family $(\mathcal{G},\Pi,\Psi,p(0))$ is exponentially mean square stable, and all members share a common uniform decay rate.

The goal in this section is to establish a necessary and sufficient condition for uniform stability that is more tractable than an infinite set of LMIs. It is well-known that any stable discrete-time linear time-varying system admits a time-varying quadratic Lyapunov function; it is less well-known that the usual construction (e.g., see~\cite[Thm. 23.3]{lutzStilwell:rugh1996book}) can be modified so that at each time instant, the Lyapunov function depends on only a finite number of the past system parameter matrices~\cite[Lem. 4]{lutzStilwell:leeDullerud2006Stab}. Inspired by this fact, the following lemma constructs a stochastic Lyapunov function for a stable time-inhomogeneous Markov jump linear system that depends on only a finite number of the future transition probability matrices.

\begin{lem}\label{lem:finiteLyapTih}
Suppose system $(\mathcal{G},\Pi,\Psi,p(0))$ is uniformly exponentially mean square stable with stability constant $c$ and decay rate $\lambda$ in Definition~\ref{defn:uemss}. Let $M=\max(\left\lceil\frac{\log(1/c)}{\log(\lambda)}-2 \right\rceil, 0)$ so that $c\lambda^{M+2}<1$. Then for each $\psi\in\Psi$, $Y_{\psi}(i,k) :=\sum_{j=k}^{k+M+1}\mathbf{E} [ \Phi^{\tpose}(j,k) \Phi(j,k) \mid \theta(k)=i ]$ satisfies
\begin{IEEEeqnarray}{rCl}
\IEEEyesnumber \label{eq:finiteLyapTihIneq}
\eta I \leq Y_{\psi}(i,k) &\leq & \rho I \IEEEyessubnumber \label{eq:finiteLyapTihIneqA}  \\
A^{\tpose}(i) \tilde{Y}_{\psi}(i,k+1) A(i) - Y_{\psi}(i,k) &\leq & -\nu I. \IEEEyessubnumber \label{eq:finiteLyapTihIneqB} \IEEEeqnarraynumspace
\end{IEEEeqnarray}
for all $i\in\mathcal{N}$ and all $k\in\mathbb{N}_0$ where $\tilde{Y}_{\psi}(i,k+1)=\sum_{j=1}^N \pi_{ij}(\psi(k+1))Y_{\psi}(j,k+1)$, $\eta=1$, $\rho=c/(1-\lambda)$, and $\nu=1-c\lambda^{M+2}$.
\end{lem}
\begin{IEEEproof}
The inequalities in~\eqref{eq:finiteLyapTihIneqA} follow readily from Definition~\ref{defn:uemss} and the definition of $Y_{\psi}$. For convenience, define $\Gamma(j,k)=\Phi^{\tpose}(j,k) \Phi(j,k)$. Then
\begin{IEEEeqnarray*}{rCl}
\IEEEeqnarraymulticol{3}{l}{
A^{\tpose}(i) \tilde{Y}_{\psi}(i,k+1) A(i)
}
\\
\quad
&=&\sum_{j=k+1}^{k+M+2}\expec{ A^{\tpose}(\theta(k)) \Gamma(j,k+1) A(\theta(k)) \mid \theta(k)=i } \IEEEyesnumber \IEEEeqnarraynumspace \label{eq:interchSumIterExpec}\\
&=& Y_{\psi}(i,k) - I + \expec{ \Gamma(k+M+2,k) \mid \theta(k)=i } \IEEEyesnumber \label{eq:finiteEq},
\end{IEEEeqnarray*}
where~\eqref{eq:interchSumIterExpec} follows by plugging in the definition of $\tilde{Y}_{\psi}(i,k+1)$, interchanging the order of summation, and recognizing an iterated expectation. Definition~\ref{defn:uemss} and equation~\eqref{eq:finiteEq} show~\eqref{eq:finiteLyapTihIneqB} with $\nu>0$ by the hypothesis on $M$.
\end{IEEEproof}

\begin{rem}\label{rem:finiteLyapTihDependsFinitePsi}
Note that $\sum_{j=k}^{k+M+1} \Phi^{\tpose}(j,k) \Phi(j,k)$ from Lemma~\ref{lem:finiteLyapTih} is a function of the random variables $(\theta(k),\dots,\theta(k+M))$. The joint probability distribution $\mathbf{P} \{ \theta(k+1)=i_{1},\dots,\theta(k+M) = i_{M} \mid \theta(k) = i_{0} \} = \prod_{l=1}^{M} \pi_{i_{l-1}i_{l}}(\psi(k+l))$ is required to compute the expectation in the definition of $Y_{\psi}(i,k)$ in Lemma~\ref{lem:finiteLyapTih}. Since the joint probability distribution is determined by the conditional value of $\theta(k)$ and the value of $\psi_{M}(k)$, $Y_{\psi}(i,k)$ may be computed with knowledge of only $i$ and $\psi_{M}(k)$. Since $\mathcal{N}$ and $\mathcal{J}$ are finite sets, $\cup_{\psi\in\Psi} \im Y_{\psi}$ is a finite set of matrices with no more than $NJ^M$ elements. Fix $\psi\in\Psi$ arbitrarily. For $i\in\mathcal{N}$, $k\in\mathbb{N}_0$, and $y\in\mathbb{R}^n$, define $V_{\psi}(i,k,y) := y^{\tpose} Y_{\psi}(i,k) y$. By~\eqref{eq:finiteLyapTihIneq}, $V_{\psi}$ is a quadratic stochastic Lyapunov function for system $(\mathcal{G},\Pi \circ \psi,p(0))$. Thus, uniform stability of the family $(\mathcal{G},\Pi,\Psi,p(0))$ guarantees the existence of a finite set of matrices that may be used to construct a time-varying quadratic stochastic Lyapunov function for any member of the family.
\end{rem}

The next theorem, inspired by~\cite[Thm. 9]{lutzStilwell:leeDullerud2006Stab}, provides a necessary and sufficient condition, expressed as a set of finite-dimensional LMIs, for uniform exponential mean square stability of a switched Markov jump linear system.
\begin{thm}\label{thm:smjUEMSS}
The switched Markov jump linear system $(\mathcal{G},\Pi,\Psi,p(0))$ is uniformly exponentially mean square stable if and only if there exist $M\in\mathbb{N}_0$ and a function $X:\mathcal{N}\times \Psi_{M} \to \mathbb{S}_{n}^{+}$ such that 
\begin{IEEEeqnarray}{c}
A^{\tpose}(i)  \sum_{j=1}^N \pi_{ij}(r_1) X(j,r_2,\dots,r_{M+1}) A(i) - X(i,r_1,\dots,r_{M}) < 0 \nonumber\\* \label{eq:smjUEMSS}
\end{IEEEeqnarray}
for any $(r_1,\dots,r_{M+1})\in\Psi_{M+1}$ and $i\in\mathcal{N}$.
\end{thm}
\begin{IEEEproof}
Suppose there exist $M$ and $X$ such that~\eqref{eq:smjUEMSS} holds. Let $\psi\in\Psi$ be arbitrary. Define $Y_{\psi}(i,k):=X(i,\psi_{M}(k))$. Since $\mathcal{N}\times\Psi_{M+1}\subset \mathcal{N}\times\mathcal{J}^{M+1}$ is a finite set, inequality~\eqref{eq:smjUEMSS} holds uniformly, and we can find $\eta,\rho,\nu>0$ such that $\eta I \leq Y_{\psi}(i,k) \leq \rho I$ and $A^{\tpose}(i) \tilde{Y}_{\psi}(i,k+1) A(i) - Y_{\psi}(i,k) \leq  -\nu I$ for all $i\in\mathcal{N}$, $k\in\mathbb{N}_0$, and $\psi\in\Psi$. Thus, $y^{\tpose}Y_{\psi}(i,k)y$ is a stochastic Lyapunov function for the single system $(\mathcal{G},\Pi \circ \psi,p(0))$ and guarantees exponential mean square stability by Proposition~\ref{prop:lyapTih} with $c=\rho/\eta$ and $\lambda=1-\nu/\rho$. Since $\psi$ was arbitrary and the same $c$ and $\lambda$ work for any $\psi\in\Psi$, $(\mathcal{G},\Pi,\Psi,p(0))$ is uniformly exponentially mean square stable.

Conversely, assume that $(\mathcal{G},\Pi,\Psi,p(0))$ is uniformly exponentially mean square stable with stability constant $c$ and decay rate $\lambda$. Fix $M\in\mathbb{N}_0$ such that $c\lambda^{M+2}<1$. Let $(i,r_1,\dots,r_{M+1})\in\mathcal{N}\times\Psi_{M+1}$ be arbitrary. By definition of $\Psi_{M+1}$, there exist $\psi\in\Psi$ and $t\in\mathbb{N}_0$ such that $\psi_{M+1}(t) = (r_1,\dots,r_{M+1})$. Construct $Y_{\psi}$ as in Lemma~\ref{lem:finiteLyapTih} and recall from Remark~\ref{rem:finiteLyapTihDependsFinitePsi} that $Y_{\psi}(i,t)$ depends only on $(i,\psi_{M}(t))$. Thus, define $X(i,r_1,\dots,r_M):= Y_{\psi}(i,t)$ and define $X(i,r_2,\dots,r_{M+1}):=Y_{\psi}(i,t+1)$. One recovers every inequality in~\eqref{eq:smjUEMSS} from~\eqref{eq:finiteLyapTihIneq}.
\end{IEEEproof}

\begin{rem}
For any $M\in\mathbb{N}_0$, $\im X$ in Theorem~\ref{thm:smjUEMSS} is finite. Thus, for each $M\in\mathbb{N}_0$, the number of LMIs specified in~\eqref{eq:smjUEMSS} is finite. The stability of a switched Markov jump linear system may be investigated using an iterative algorithm. First, set $M=0$ and check if the LMIs in~\eqref{eq:smjUEMSS} are feasible. If not, increment $M$ and repeat. If the switched Markov jump linear system is stable, Theorem~\ref{thm:smjUEMSS} says that this algorithm will stop in a finite amount of time with some finite value of $M$. A conservative estimate for $M$ is based on the uniform decay rate of the switched Markov jump linear system (see Lemma~\ref{lem:finiteLyapTih}).
\end{rem}

\begin{rem}
Theorem~\ref{thm:smjUEMSS} provides a practical approach for investigating the stability of a \emph{single} time-inhomogeneous Markov jump linear system with \emph{known} transition probability matrices that vary in a finite set (let $\Psi$ be the set containing a single sequence).
\end{rem}

\begin{rem}\label{rem:stabReducesToThWhenSingleton}
Consider the case when $J=1$ and $\Psi = \{ (1,1,\dots) \}$. The switched Markov jump linear system $(\mathcal{G},\Pi,\Psi,p(0))$ reduces to a single time-homogeneous Markov jump linear system $(\mathcal{G},\Pi \circ \psi_1,p(0))$ where $\psi_1\equiv 1$. For any $M$, the set $\Psi_{M}$ contains only a single element $(1,\dots,1)$, and the set $\mathcal{N}\times\Psi_{M}$ contains only $N$ elements. For $i\in\mathcal{N}$, define $Z(i):=X(i,1,\dots, 1)$ where $X$ is as in Theorem~\ref{thm:smjUEMSS}. Then~\eqref{eq:smjUEMSS} reduces to $A^{\tpose}(i) \sum_{j=1}^N \pi_{ij}(1) Z(j) A(i) - Z(i) < 0$, which is the well-known stability criterion (see~\cite[Thm. 2.1]{lutzStilwell:ji1990a} or~\cite[Thm. 2]{lutzStilwell:costa1993}) for time-homogeneous Markov jump linear systems; this well-known result is a corollary of Theorem~\ref{thm:smjUEMSS}.
\end{rem}

\subsection{Disturbance Attenuation}
We now address disturbance attenuation for a time-inhomogeneous Markov jump linear system where the sequence of transition probability matrices is not known a priori. Accordingly, we modify the definition of disturbance attenuation so that it applies uniformly over all possible sequences of transition probability matrices.
\begin{defn}\label{defn:umssc}
The switched Markov jump linear system $(\mathcal{G},\Pi,\Psi,p(0))$ is \emph{uniformly mean square strictly contractive} if there exists $\gamma\in(0,1)$ such that whenever $x(0)=0$, $\norm{z}_{2,e} \leq \gamma \norm{w}_{2,e}$ for all $w\in\ell^2_e$ and all $\psi\in\Psi$.
\end{defn}

The goal of this section is to establish a KYP-like result for switched Markov jump linear systems in terms of finite-dimensional LMIs similar to Theorem~\ref{thm:smjUEMSS}. The main result can be found in Theorem~\ref{thm:smjlsUMSSC}. The necessity of the LMIs is the difficult part of the proof and hinges on the existence of the matrix-valued functions in Lemma~\ref{lem:FinDependRicSolnTeaser}. Like Lemma~\ref{lem:finiteLyapTih}, at any time instant each matrix-valued function depends only on a finite number of the future transition probability matrices.
\begin{lem}\label{lem:FinDependRicSolnTeaser}
Assume $p_i(k)>0$ for all $\psi\in\Psi$, $i\in\mathcal{N}$, and $k\in\mathbb{N}_0$. If $(\mathcal{G},\Pi,\Psi,p(0))$ is uniformly exponentially mean square stable and uniformly mean square strictly contractive then there exist $\eta,\rho,\nu>0$ and $M\in\mathbb{N}_0$ such that for each $\psi\in\Psi$, there exists $Y_\psi:\mathcal{N}\times\mathbb{N}_0\to\mathbb{S}_{n}^{+}$ such that $Y_{\psi}(i,k)$ depends only on $i$ and $\psi_{M}(k)$ and satisfies
\begin{IEEEeqnarray}{rCl}
\IEEEyesnumber \label{eq:FinDependRicSolnTeaser}
\eta I  \leq  Y_{\psi}(i,k) & \leq & \rho I \IEEEyessubnumber \label{eq:FinDependRicSolnTeaserA}
\\
\mathcal{B}(i,\tilde{Y}_{\psi}(i,k+1),Y_{\psi}(i,k))
& \leq & -\nu I
\IEEEyessubnumber \IEEEeqnarraynumspace \label{eq:FinDependRicSolnTeaserB}
\end{IEEEeqnarray}
for all $i\in\mathcal{N}$ and all $k\in\mathbb{N}_0$ where $\tilde{Y}_{\psi}(i,k+1) = \sum_{j=1}^N \pi_{ij}(\psi(k+1)) Y_{\psi}(j,k+1)$.
\end{lem}

The construction of the functions $Y_{\psi}$, $\psi\in\Psi$ requires the intermediate results contained in this section up to and including Lemma~\ref{lem:cEpsLambdaEpsExist}, and the proof of Lemma~\ref{lem:FinDependRicSolnTeaser} follows directly from Lemma~\ref{lem:smjlsRiccatiFiniteDependence}. For the moment, suppose that functions $Y_{\psi}$, $\psi\in\Psi$ have been found that satisfy Lemma~\ref{lem:FinDependRicSolnTeaser} and define $V_{\psi}(i,k,y) := y^{\tpose} Y_{\psi}(i,k) y$ for $i\in\mathcal{N}$, $k\in\mathbb{N}_0$, and $y\in\mathbb{R}^n$. Then $V_{\psi}$ is a quadratic stochastic storage function for system $(\mathcal{G},\Pi\circ \psi,p(0))$ that at each time instant depends only on $i$ and $\psi_{M}(k)$. Since $\mathcal{N}$ and $\mathcal{J}$ are finite sets, $\cup_{\psi\in\Psi}\im Y_{\psi}$ is a finite set of matrices with no more than $NJ^M$ elements. Thus, uniform stability and contractiveness of the switched Markov jump linear system $(\mathcal{G},\Pi,\Psi,p(0))$ guarantees the existence of a finite set of matrices that may be used to construct a time-varying quadratic stochastic storage function for any individual Markov jump linear system in the family.

Riccati difference equations defined in terms of the following operators play a key role in the construction of the functions $Y_{\psi}$, $\psi\in\Psi$ in Lemma~\ref{lem:FinDependRicSolnTeaser}.
\begin{defn}\label{def:riccatiOperators}
Let $\mathcal{G}$ be given. For $i\in\mathcal{N}$ and $X\in\mathbb{S}_{n}$, define 
\begin{IEEEeqnarray}{rCl}
\mathcal{L}(i,X) &=& A^{\tpose}(i)X A(i) + C^{\tpose}(i)C(i) \nonumber \\
\mathcal{R}(i,X) &=& B^{\tpose}(i)XA(i) + D^{\tpose}(i)C(i) \nonumber \\
\mathcal{W}(i,X) &=& I - B^{\tpose}(i) X B(i) - D^{\tpose}(i)D(i) \nonumber \\
\mathcal{M}(i,X) &=& 
\begin{bmatrix}
\mathcal{L}(i,X) & \mathcal{R}^{\tpose}(i,X) \\
\mathcal{R}(i,X) & -\mathcal{W}(i,X)
\end{bmatrix} \nonumber
\end{IEEEeqnarray}
For $i\in\mathcal{N}$ let $\mathbb{X}_i = \{ X\in\mathbb{S}_{n} : \mathcal{W}(i,X) \text{ invertible} \}$. For $i\in\mathcal{N}$ and $X\in\mathbb{X}_i$, define
\begin{IEEEeqnarray}{rCl}
\mathcal{S}(i,X) &=& \mathcal{L}(i,X) + \mathcal{R}^{\tpose}(i,X)\mathcal{W}^{-1}(i,X)\mathcal{R}(i,X). \nonumber
\end{IEEEeqnarray}
Given a modified set of matrices $\{(A(i),B(i),C_\epsilon(i),D_\epsilon(i)) : i\in\mathcal{N}\}$, let $\mathcal{L}_\epsilon(i,X)$, $\mathcal{R}_\epsilon(i,X)$, $\mathcal{W}_\epsilon(i,X)$, and $\mathcal{S}_\epsilon(i,X)$ be defined as above but with $C_\epsilon(i)$ in place of $C(i)$ and $D_\epsilon(i)$ in place of $D(i)$.
\end{defn}

Note that inequality~\eqref{eq:FinDependRicSolnTeaserB} may be rewritten in terms of the operators in Definition~\ref{def:riccatiOperators}. Expanding the left side of~\eqref{eq:FinDependRicSolnTeaserB} gives
\begin{IEEEeqnarray}{rCl}
\IEEEeqnarraymulticol{3}{l}{
\mathcal{B}(i,\tilde{Y}_{\psi}(i,k+1),Y_{\psi}(i,k))
} \nonumber
\\
\quad &=&
\mathcal{M}(i,\tilde{Y}_{\psi}(i,k+1)) - 
\begin{bmatrix}
Y_{\psi}(i,k) & 0 \\
0 & 0
\end{bmatrix}. \IEEEeqnarraynumspace \label{eq:MjlsTihBrlBDiffForm}
\end{IEEEeqnarray} 
By the Schur complement,~\eqref{eq:MjlsTihBrlBDiffForm} is negative definite if and only if $\mathcal{W}(i,\tilde{Y}_{\psi}(i,k+1)) > 0$ and $Y_{\psi}(i,k) > \mathcal{S}(i,\tilde{Y}_{\psi}(i,k+1))$. Using these inequalities as a guide, we shall examine finite-horizon Riccati difference equations defined by the recursive relation and final condition
\begin{IEEEeqnarray}{rCl}
\IEEEyesnumber \label{eq:smjlsRiccatiFinHorizRecurAndFinalCond}
X_{\psi}(i,k,T) &=& \mathcal{S}(i,\tilde{X}_{\psi}(i,k+1,T)) \IEEEyessubnumber \label{eq:smjlsRiccatiFinHorizRecurDef}\\
X_{\psi}(i,T+1,T) &=& 0 \IEEEyessubnumber \label{eq:smjlsRiccatiFinHorizFinalCond}
\end{IEEEeqnarray}
where $i\in\mathcal{N}$, $T\in\mathbb{N}_0$ (the horizon), $0\leq k\leq T$, $\psi\in\Psi$, and $\tilde{X}_{\psi}(i,k+1,T) = \sum_{j=1}^N \pi_{ij}(\psi(k+1))X_{\psi}(j,k+1,T)$. For a fixed $\psi\in\Psi$ and $T\in\mathbb{N}_0$, the solution $X_{\psi}(\cdot,\cdot,T)$ to~\eqref{eq:smjlsRiccatiFinHorizRecurAndFinalCond} may be computed iteratively backwards-in-time starting with the final condition. However, we first need to verify that the inverse specified in~\eqref{eq:smjlsRiccatiFinHorizRecurDef} is well-defined. The algebraic identity and special input in the next lemma aid in this task.
\begin{lem}\label{lem:algIdentSpecInput}
Let $k\in\mathbb{N}_0$, $\theta(k)\in\mathcal{N}$, $X\in\mathbb{S}_{n}$, and $x(k)$, $z(k)$, $w(k)$ be as in~\eqref{eq:mjls}. Then
\begin{IEEEeqnarray}{rCl}
\IEEEeqnarraymulticol{3}{l}{
z^{\tpose}(k) z(k) - w^{\tpose}(k) w(k) + x^{\tpose}(k+1) X x(k+1)
} \nonumber
\\ \quad
&=& 
\begin{bmatrix}
x(k) \\
w(k)
\end{bmatrix}^{\tpose}
\mathcal{M}(\theta(k),X)
\begin{bmatrix}
x(k) \\
w(k)
\end{bmatrix}.
\IEEEeqnarraynumspace \label{eq:l2NormMatrixIdent}
\end{IEEEeqnarray}
If $X\in\mathbb{X}_{\theta(k)}$ and $w(k) = \mathcal{W}^{-1}(\theta(k),X)\mathcal{R}(\theta(k),X)x(k)$ then 
\begin{IEEEeqnarray}{c}
\begin{bmatrix}
x(k) \\
w(k)
\end{bmatrix}^{\tpose}
\mathcal{M}(\theta(k),X)
\begin{bmatrix}
x(k) \\
w(k)
\end{bmatrix}
=
x^{\tpose}(k) \mathcal{S}(\theta(k),X) x(k).
\label{eq:l2NormMatrixIdentSpecialwk}
\end{IEEEeqnarray}
\end{lem}
\begin{IEEEproof}
The proof of Lemma~\ref{lem:algIdentSpecInput} follows from simple matrix algebra.
\end{IEEEproof}

The following lemma establishes that the Riccati recursive relation in~\eqref{eq:smjlsRiccatiFinHorizRecurDef} is well-defined when $(\mathcal{G},\Pi,\Psi,p(0))$ is uniformly mean square strictly contractive.
\begin{lem}\label{lem:WiBoundedBelow}
Assume $p_i(k)>0$ for all $\psi\in\Psi$, $i\in\mathcal{N}$, and $k\in\mathbb{N}_0$. If $(\mathcal{G},\Pi,\Psi,p(0))$ is uniformly mean square strictly contractive then there exists $\nu>0$ such that
\begin{IEEEeqnarray*}{c}
\mathcal{W}(i,\tilde{X}_{\psi}(i,k+1,T))\geq \nu I
\end{IEEEeqnarray*}
for all $\psi\in\Psi$, $i\in\mathcal{N}$, $T\in\mathbb{N}_0$, and $0\leq k\leq T$ where $X_{\psi}$ is defined by the recursive relation and final condition in~\eqref{eq:smjlsRiccatiFinHorizRecurAndFinalCond}.
\end{lem}
\begin{IEEEproof}
Let $x(0)=0$. Arbitrarily fix $\psi\in\Psi$ and $T\in\mathbb{N}_0$. Let $i\in\mathcal{N}$ and consider $w$ such that $w(T) = \chi\{\theta(T) = i\} y$, and $w(k) = 0$ for $k\neq T$, where $y$ is an arbitrary vector, and $\chi\{\theta(T) = i\}$ is the indicator function of the set $\{\theta(T)=i\}\subset\Omega$. Note that $\expec{\chi\{\theta(T) = i\}}=\prob{\theta(T)=i}$. With $w$ defined above, $x(k)=0$ for $k\leq T$ and $z(k) = 0$ for $k\leq T-1$. Definition~\ref{defn:umssc} gives 
\begin{IEEEeqnarray}{c}
\norm{z}_{2,e}^2 - \norm{w}_{2,e}^2 \leq -\nu \norm{w}_{2,e}^2 \label{eq:contractiveDefSubw}
\end{IEEEeqnarray}
for any $\psi\in\Psi$ and $w\in\ell^2_e$ where $\nu = 1 - \gamma^2$. Then
\begin{IEEEeqnarray}{rCl}
\IEEEeqnarraymulticol{3}{l}{
-\nu\norm{w}_{2,e}^2  =  -\nu \prob{\theta(T)=i} y^{\tpose}y
}\label{eq:expecwT}\\ \quad \quad \quad
& \geq & \sum_{k=0}^T \expec{ \norm{z(k)}^2 - \norm{w(k)}^2 } \label{eq:sumZeroTineqNu} \\
& = & \mathbf{E} \big[ z^{\tpose}(T) z(T) - w^{\tpose}(T)w(T) \nonumber \\
&& \quad +\, x^{\tpose}(T+1) \tilde{X}_{\psi}(\theta(T),T+1,T) x(T+1) \big] \IEEEeqnarraynumspace \label{eq:finalConditionZero} \\
& = & \expec{ -w^{\tpose}(T)\mathcal{W}(\theta(T),\tilde{X}_{\psi}(\theta(T),T+1,T)) w(T) } \IEEEeqnarraynumspace \label{eq:xTisZero} \\
& = & - \prob{\theta(T) = i} y^{\tpose} \mathcal{W}(i,\tilde{X}_{\psi}(i,T+1,T)) y \nonumber
\end{IEEEeqnarray}
where~\eqref{eq:expecwT} follows from the definition of $w$; \eqref{eq:sumZeroTineqNu} follows from~\eqref{eq:contractiveDefSubw}; \eqref{eq:finalConditionZero} follows since $X_{\psi}(i,T+1,T)=0$; and,~\eqref{eq:xTisZero} follows from~\eqref{eq:l2NormMatrixIdent} and $x(T)=0$. Thus, $\mathcal{W}(i,\tilde{X}_{\psi}(i,T+1,T))\geq \nu I$ since $y$ was an arbitrary vector.

Now fix $0\leq t\leq T$ and assume $\mathcal{W}(i,\tilde{X}_{\psi}(i,k+1,T))\geq \nu I$ for $t\leq k \leq T$ and $i\in\mathcal{N}$. Consider $w$ of the form
\begin{IEEEeqnarray}{c}
w(k) = \left\{ \,
     \begin{IEEEeqnarraybox}[][c]{l.C.l}
       \IEEEstrut
       0 & : & k\leq t-2 \\
       \chi\{\theta(k)=i\} y & : & k = t-1 \\
       \mathcal{W}^{-1}(\theta(k),\tilde{X}_{\psi}(\theta(k),k+1,T)) \\ \; \times \mathcal{R}(\theta(k),\tilde{X}_{\psi}(\theta(k),k+1,T))x(k) & : & t\leq k\leq T \\
       0 & : & k\geq T+1
       \IEEEstrut
     \end{IEEEeqnarraybox}
   \right. . \IEEEeqnarraynumspace \label{eq:specialwIndicFuncAndSchurTerms}
\end{IEEEeqnarray}
Then $x(k)=0$ for $k\leq t-1$ and $z(k)=0$ for $k\leq t-2$. Define $\mathcal{V}(k) = x^{\tpose}(k) X_{\psi}(\theta(k),k,T) x(k)$ and $\tilde{\mathcal{V}}(k) = x^{\tpose}(k) \tilde{X}_{\psi}(\theta(k-1),k,T) x(k)$. Then
\begin{IEEEeqnarray}{rCl}
\IEEEeqnarraymulticol{3}{l}{
\sum_{k=0}^T \expec{ \norm{z(k)}^2 - \norm{w(k)}^2 } 
} \label{eq:sumZeroT} \\ \; 
&=& \sum_{k=t-1}^T \expec{ \norm{z(k)}^2 - \norm{w(k)}^2 + \tilde{\mathcal{V}}(k+1) - \mathcal{V}(k) } \label{eq:iterExpecAndVT1zero} \\
&=& \expec{ -w^{\tpose}(t-1) \mathcal{W}(\theta(t-1),\tilde{X}_{\psi}(\theta(t-1),t,T)) w(t-1) } \IEEEeqnarraynumspace \label{eq:optimalwkcancels} \\
&=& - \prob{\theta(t-1)=i} y^{\tpose} \mathcal{W}(i,\tilde{X}_{\psi}(i,t,T))y \nonumber \\
&\leq & - \nu \expec{ w^{\tpose}(t-1)w(t-1) } \label{eq:applySumZeroTIneqNu} \\
&=& -\nu\prob{\theta(t-1)=i} y^{\tpose}y \nonumber
\end{IEEEeqnarray}
where~\eqref{eq:iterExpecAndVT1zero} follows after recognizing a telescoping sum, realizing $\mathcal{V}(t-1)=\mathcal{V}(T+1)=0$, and applying an iterated expectation; \eqref{eq:optimalwkcancels} follows from Lemma~\ref{lem:algIdentSpecInput}, $x(t-1)=0$, the recursive relation~\eqref{eq:smjlsRiccatiFinHorizRecurDef}, and definition of $\mathcal{V}(k)$; and,~\eqref{eq:applySumZeroTIneqNu} follows from~\eqref{eq:contractiveDefSubw}. Since $y$ was an arbitrary vector, $\mathcal{W}(i,\tilde{X}_{\psi}(i,t,T))\geq \nu I$. The result follows by induction.
\end{IEEEproof}

\begin{rem}\label{rmk:RiccatiX0Rmk}
The input specified in~\eqref{eq:specialwIndicFuncAndSchurTerms} is similar to disturbance inputs constructed in~\cite{lutzStilwell:seiler2003} and~\cite{lutzStilwell:aberkane2013}. The techniques used in Lemma~\ref{lem:WiBoundedBelow} show that if
\begin{IEEEeqnarray}{c}
w(k) = \left\{ \,
     \begin{IEEEeqnarraybox}[][c]{lCl}
       \IEEEstrut
       \mathcal{W}^{-1}(\theta(k),\tilde{X}_{\psi}(\theta(k),k+1,T)) \\ \; \times \mathcal{R}(\theta(k),\tilde{X}_{\psi}(\theta(k),k+1,T))x(k) & : & 0\leq k\leq T \\
       0 & : & k\geq T+1
       \IEEEstrut
     \end{IEEEeqnarraybox}
   \right. \IEEEeqnarraynumspace \label{eq:inputToGetRiccatiX0}
\end{IEEEeqnarray}
then $\sum_{k=0}^T \expec{z^{\tpose}(k)z(k) - w^{\tpose}(k)w(k)} = \expec{x^{\tpose}(0) X_{\psi}(\theta(0),0,T) x(0)}$. The disturbance input~\eqref{eq:inputToGetRiccatiX0} maximizes the quantity in~\eqref{eq:sumZeroT} (see~\cite[Lem. 2.1]{lutzStilwell:limebeer1989}).
\end{rem}

A hypothesis in the statement of Lemma~\ref{lem:WiBoundedBelow} can be expressed as a requirement on the possible sequences $\Pi(\psi(k))$ of stochastic matrices and the initial distribution $p(0)$.
\begin{prop}\label{prop:probHypToInitDistAndPkHyp}
For all $\psi\in\Psi$, any $i\in\mathcal{N}$, and all $k\in\mathbb{N}_0$, $p_i(k)>0$ if and only if for all $\psi\in\Psi$ and all $k\in\mathbb{N}$, each column of $\Pi(\psi(k))$ is nonzero, and $p_i(0)>0$ for all $i\in\mathcal{N}$.
\end{prop}
\begin{IEEEproof}
Use induction and the identity $\mathbf{P} \{ \theta(k) = i \} = \sum_{l=1}^N \pi_{li}(\psi(k))\mathbf{P} \{ \theta(k-1) = l \}$.
\end{IEEEproof}

The following property is key for finding a uniform upper bound on solutions to~\eqref{eq:smjlsRiccatiFinHorizRecurAndFinalCond}.
\begin{lem}\label{lem:timeInvarianceLikeRiccati}
Fix $\psi\in\Psi$ and $t\in\mathbb{N}_0$. Define $\psi_t$ to be a shifted version of $\psi$ so that $\psi_{t}(k)=\psi(t+k)$ for $k\in\mathbb{N}$, and define $p_{t}(0)=p(t)$. If $(\mathcal{G},\Pi\circ\psi,p(0))$ is exponentially mean square stable and mean square strictly contractive, then $(\mathcal{G},\Pi\circ\psi_{t},p_{t}(0))$ is exponentially mean square stable and mean square strictly contractive. Furthermore, 
\begin{IEEEeqnarray}{c}
X_{\psi}(i,t,T) = X_{\psi_{t}}(i,0,T-t) \label{eq:timeInvarianceLikeRiccatiClaim}
\end{IEEEeqnarray}
for $i\in\mathcal{N}$ and $0\leq t \leq T$ where $X_{\psi}$ and $X_{\psi_{t}}$ are defined by~\eqref{eq:smjlsRiccatiFinHorizRecurAndFinalCond}.
\end{lem}
\begin{IEEEproof}
Consider the Markov jump linear system modulated by a shifted random process
\begin{IEEEeqnarray}{c}\label{eq:shiftedMjls}
\begin{bmatrix}
x_{t}(k+1) \\
z_{t}(k)
\end{bmatrix}
= 
\begin{bmatrix}
A(\theta_{t}(k)) & B(\theta_{t}(k)) \\
C(\theta_{t}(k)) & D(\theta_{t}(k))
\end{bmatrix}
\begin{bmatrix}
x_{t}(k) \\
w_{t}(k)
\end{bmatrix}
\end{IEEEeqnarray}
where $\theta_{t}(k) = \theta(t+k)$ for $k\in\mathbb{N}_0$ and $w_{t}\in\ell_{e}^{2}$. Note that this system may be denoted by $(\mathcal{G},\Pi\circ\psi_{t},p_{t}(0))$. Now $(\mathcal{G},\Pi\circ\psi_{t},p_{t}(0))$ is exponentially mean square stable since $\mathbf{E}\big[ \Phi_{t}^{\tpose}(k,j) \Phi_{t}(k,j) \mid \theta_{t}(j) \big] = \mathbf{E}\big[ \Phi^{\tpose}(t+k,t+j) \Phi(t+k,t+j) \mid \theta(t+j) \big] \leq c \lambda^{t+k - (t+j)} I$ where $\Phi_{t}$ is the random state transition matrix for the system in~\eqref{eq:shiftedMjls}. Now let $x_t(0)=x(0)=0$, and let $w_t\in\ell_{e}^{2}$ be arbitrary. Define $w$ such that $w(k) = 0$ when $k<t$, and $w(k) = w_{t}(k-t)$ when $k\geq t$. Note that $w_t\in\ell_{e}^{2}$ implies $w\in\ell_{e}^{2}$, and $\norm{w}_{2,e}=\norm{w_{t}}_{2,e}$. Furthermore, $z(k)=0$ for $0\leq k \leq t-1$, and $x(k)=0$ for $0\leq k \leq t$. It is easily shown that $z_{t}(k)=z(t+k)$ for $k\in\mathbb{N}_0$ and $\norm{z_{t}}_{2,e}=\norm{ z }_{2,e}$. Since $(\mathcal{G},\Pi\circ\psi,p(0))$ is mean square strictly contractive, $\norm{z_{t}}_{2,e}=\norm{ z }_{2,e} \leq \gamma \norm{w}_{2,e} = \gamma \norm{w_{t}}_{2,e}$. Since $w_t\in\ell_{e}^{2}$ was arbitrary, $(\mathcal{G},P_{t},p_{t}(0))$ is mean square strictly contractive.

Now to prove~\eqref{eq:timeInvarianceLikeRiccatiClaim}, note that the base case $X_{\psi}(i,T,T)=X_{\psi_{t}}(i,T-t,T-t)=\mathcal{S}(i,0)$ holds for all $i\in\mathcal{N}$. For the inductive hypothesis, assume for some $0\leq k \leq T-t-1$ that $X_{\psi}(i,T-k,T)=X_{\psi_{t}}(i,T-t-k,T-t)$ for all $i\in\mathcal{N}$. Then
\begin{IEEEeqnarray}{rCl}
\IEEEeqnarraymulticol{3}{l}{
X_{\psi}(i,T-(k+1),T)
} \nonumber \\ \quad
&=& \mathcal{S}\left(i,\sum_{j=1}^N \pi_{ij}(\psi(T-k)) X_{\psi}(j,T-k,T)\right)
\nonumber \\
&=& \mathcal{S}\left(i,\sum_{j=1}^N \pi_{ij}(\psi_{t}(T-t-k)) X_{\psi_{t}}(j,T-t-k,T-t)\right) \IEEEeqnarraynumspace \label{eq:psiSigmaEqual}
\\
&=& X_{\psi_{t}}(i,T-t-(k+1),T-t) \nonumber
\end{IEEEeqnarray}
where~\eqref{eq:psiSigmaEqual} follows from the inductive hypothesis and the fact that $\psi(T-k)=\psi_{t}(T-t-k)$. Equation~\eqref{eq:timeInvarianceLikeRiccatiClaim} follows by induction.
\end{IEEEproof}

A uniform upper bound on solutions to~\eqref{eq:smjlsRiccatiFinHorizRecurAndFinalCond} is established in the following lemma using Lemma~\ref{lem:timeInvarianceLikeRiccati} and a technique similar to~\cite[Sec. B.2.3]{lutzStilwell:greenLimebeer1995}.
\begin{lem}\label{lem:RiccatiXkBoundedAboveUnif}
Assume $p_i(k)>0$ for all $\psi\in\Psi$, $i\in\mathcal{N}$, and $k\in\mathbb{N}_0$. If $(\mathcal{G},\Pi,\Psi,p(0))$ is uniformly exponentially mean square stable and uniformly mean square strictly contractive then there exists $\rho>0$ such that 
\begin{IEEEeqnarray}{c}
0\leq X_{\psi}(i,k,T)\leq \rho I \label{eq:RiccatiXkBoundedAboveUnif}
\end{IEEEeqnarray}
for all $i\in\mathcal{N}$, any $T\in\mathbb{N}_0$, all $\psi\in\Psi$, and all $0\leq k \leq T+1$ where $X_{\psi}$ is defined in~\eqref{eq:smjlsRiccatiFinHorizRecurAndFinalCond}.
\end{lem}
\begin{IEEEproof}
Arbitrarily fix $\psi\in\Psi$ and $T\in\mathbb{N}_0$. Define $w$ as in~\eqref{eq:inputToGetRiccatiX0}. Then
\begin{IEEEeqnarray}{rCl}
\IEEEeqnarraymulticol{3}{l}{
\expec{x^{\tpose}(0) X_{\psi}(\theta(0),0,T) x(0)}
}
\nonumber \\ \quad
&=& \sum_{k=0}^T\expec{ z^{\tpose}(k)z(k) - w^{\tpose}(k)w(k) } \IEEEeqnarraynumspace \label{eq:RiccatiX0Rmk}\\
&\leq & \norm{z}_{2,e}^2 - \norm{w}_{2,e}^2 \IEEEeqnarraynumspace \label{eq:RiccatiX0ineqzw}
\end{IEEEeqnarray}
where~\eqref{eq:RiccatiX0Rmk} follows from Remark~\ref{rmk:RiccatiX0Rmk}. By linearity, $z = z_{x_0} + z_{w}$ where $z_{x_0}$ is the zero-input response and $z_{w}$ is the zero-state response (e.g., see~\cite[Ch. 20]{lutzStilwell:rugh1996book}). By the Cauchy-Schwarz inequality
\begin{IEEEeqnarray}{c}\label{eq:zCauchySchwarz}
\norm{z}_{2,e}^2 \leq \norm{z_{w}}_{2,e}^2 + \norm{z_{x_0}}_{2,e}^2 + 2 \norm{z_w}_{2,e} \norm{z_{x_0}}_{2,e}.
\end{IEEEeqnarray}
Since $(\mathcal{G},\Pi,\Psi,p(0))$ is uniformly exponentially mean square stable,
\begin{IEEEeqnarray}{rCl}
\norm{z_{x_0}}_{2,e}^2 &=& \sum_{k=0}^\infty \expec{ x^{\tpose}(0) \Phi^{\tpose}(k,0) C^{\tpose}(\theta(k)) C(\theta(k)) \Phi(k,0) x(0) } \nonumber\\
&\leq & \max_{i\in\mathcal{N}} \left( \lambda_{\text{max}}( C^{\tpose}(i)C(i)) \right) \sum_{k=0}^\infty c \lambda^k \expec{ \norm{x(0)}^2 } \nonumber\\
&=& \delta \expec{ \norm{x(0)}^2 } \label{eq:deltaBound}
\end{IEEEeqnarray}
for all $\psi\in\Psi$ where $\delta=\max_{i\in\mathcal{N}} \left( \lambda_{\text{max}}( C^{\tpose}(i)C(i)) \right)c / (1-\lambda)$ and $\lambda_{\text{max}}(\cdot)$ denotes the maximum eigenvalue. By Definition~\ref{defn:umssc},
\begin{IEEEeqnarray}{rCl}
\norm{z_{w}}_{2,e}^2 -  \norm{w}_{2,e}^2 &\leq & -\nu\norm{w}_{2,e}^2 \label{eq:zeroStateContractive1} \\
\norm{z_{w}}_{2,e} &<& \norm{w}_{2,e} \label{eq:zeroStateContractive2}
\end{IEEEeqnarray}
for all $\psi\in\Psi$ and $w\in\ell^2_e$ where $\nu=1-\gamma^2$. Then
\begin{IEEEeqnarray}{rCl}
\IEEEeqnarraymulticol{3}{l}{
\norm{z}_{2,e}^2 - \norm{w}_{2,e}^2
} \nonumber \\
&\leq & -\nu\norm{w}_{2,e}^2 + \delta\mathbf{E}[\, \lVert x(0)\rVert^2 \,] 
+ 2  \sqrt{\delta\mathbf{E}[\, \lVert x(0)\rVert^2 \,]} \norm{w}_{2,e} \label{eq:zMinusWcombineIneqs} \\
&=& \left(\delta + \delta/\nu \right)\mathbf{E}[\, \lVert x(0)\rVert^2 \,] \nonumber \\
&& \quad -\, \nu \left( \norm{w}_{2,e} - \sqrt{\delta} / \nu \sqrt{\mathbf{E}[\, \lVert x(0)\rVert^2 \,]} \right)^2 \IEEEeqnarraynumspace \label{eq:completeTheSquare} \\
&\leq &\rho \mathbf{E}[\, \lVert x(0)\rVert^2 \,] \label{eq:rhoBoundOnRiccatiX0}
\end{IEEEeqnarray}
for all $\psi\in\Psi$ and all $w\in\ell^2_e$ where $\rho = \delta + \delta/\nu$; \eqref{eq:zMinusWcombineIneqs} follows from~\eqref{eq:zCauchySchwarz},~\eqref{eq:deltaBound},~\eqref{eq:zeroStateContractive1}, and~\eqref{eq:zeroStateContractive2}; and,~\eqref{eq:completeTheSquare} follows by completing the square.
Choose any $i\in\mathcal{N}$ and let $x(0) = \chi\{\theta(0) = i\} y$ where $y$ is an arbitrary vector. Then~\eqref{eq:RiccatiX0ineqzw} and~\eqref{eq:rhoBoundOnRiccatiX0} imply $\mathbf{P} \{ \theta(0) = i \} y^{\tpose} X_{\psi}(i,0,T) y \leq \rho \mathbf{P} \{ \theta(0) = i \} y^{\tpose}y$. Since $y$ was arbitrary, the upper bound in~\eqref{eq:RiccatiXkBoundedAboveUnif} holds for $k=0$.
The general case follows from Lemma~\ref{lem:timeInvarianceLikeRiccati}. That $0\leq X_{\psi}(i,k,T)$ for all $i\in\mathcal{N}$, $T\in\mathbb{N}_0$, $\psi\in\Psi$, and $0\leq k \leq T+1$ can be seen clearly from~\eqref{eq:smjlsRiccatiFinHorizRecurAndFinalCond}.
\end{IEEEproof}

We now examine perturbed finite-horizon Riccati difference equations defined by the recursive relation and final condition
\begin{IEEEeqnarray}{rCl}
\IEEEyesnumber \label{eq:smjlsRiccatiExistThmRecurDefAndFinalCond}
X_{\psi}(i,k,T,\epsilon) &=& \mathcal{S}(i,\tilde{X}_{\psi}(i,k+1,T,\epsilon)) + \epsilon I \IEEEyessubnumber \label{eq:smjlsRiccatiExistThmRecurDef}\\
X_{\psi}(i,T+1,T,\epsilon) &=& 0 \IEEEyessubnumber \label{eq:smjlsRiccatiExistThmFinalCond}
\end{IEEEeqnarray}
where $i\in\mathcal{N}$, $T\in\mathbb{N}_0$, $0\leq k \leq T$, $\psi\in\Psi$, $\epsilon\geq 0$, and $\tilde{X}_{\psi}(i,k+1,T,\epsilon) = \sum_{j=1}^N \pi_{ij}(\psi(k+1))X_{\psi}(j,k+1,T,\epsilon)$. For fixed $\psi\in\Psi$, $T\in\mathbb{N}_0$, and $\epsilon$, the solution $X_\psi(\cdot,\cdot,T,\epsilon)$ to~\eqref{eq:smjlsRiccatiExistThmRecurDefAndFinalCond} may be computed iteratively backwards-in-time starting with the final condition. An augmented and perturbed system utilized in the following theorem shows that solutions to~\eqref{eq:smjlsRiccatiExistThmRecurDefAndFinalCond} are uniformly positive definite as well as uniformly bounded. 
\begin{thm}\label{thm:smjlsRiccatiExistThm}
Assume $p_i(k)>0$ for all $\psi\in\Psi$, $i\in\mathcal{N}$, and $k\in\mathbb{N}_0$. If $(\mathcal{G},\Pi,\Psi,p(0))$ is uniformly exponentially mean square stable and uniformly mean square strictly contractive then there exist $\eta,\rho,\nu>0$ such that for all $\epsilon\in [0,\eta]$, $\nu I \leq  \mathcal{W}(i,\tilde{X}_{\psi}(i,k+1,T,\epsilon))$ and
\begin{IEEEeqnarray}{rCl}
\epsilon I &\leq & X_{\psi}(i,k,T,\epsilon) \leq \rho I \label{eq:smjlsRiccatiExistThmXikPosDefDecres}
\end{IEEEeqnarray}
for all $i\in\mathcal{N}$, $T\in\mathbb{N}_0$, $0\leq k\leq T$, and $\psi\in\Psi$ where $X_{\psi}$ is defined by the recursive relation and final condition in~\eqref{eq:smjlsRiccatiExistThmRecurDefAndFinalCond}.
\end{thm}
\begin{IEEEproof}
Consider the augmented switched Markov jump linear system $(\mathcal{G}_{\epsilon},\Pi,\Psi,p(0))$ where $\mathcal{G}_{\epsilon} = \{ (A(i),B(i),C_{\epsilon}(i),D_{\epsilon}(i)) : i\in\mathcal{N} \}$ where $C_{\epsilon}(i) = [ C^{\tpose}(i) \,  \sqrt{\epsilon} I ]^{\tpose}$, and $D_{\epsilon}(i) = [ D^{\tpose}(i) \, 0 ]^{\tpose}$. First note $(\mathcal{G}_{\epsilon},\Pi,\Psi,p(0))$ is uniformly exponentially mean square stable for any $\epsilon$ since $\mathcal{G}$ and $\mathcal{G}_{\epsilon}$ share the same matrices $A(i)$, $i\in\mathcal{N}$. Since $(\mathcal{G},\Pi,\Psi,p(0))$ is uniformly mean square strictly contractive, there exists $\eta>0$ small enough so that for all $\epsilon\in[0,\eta]$ the augmented system $(\mathcal{G}_{\epsilon},\Pi,\Psi,p(0))$ is uniformly mean square strictly contractive. By Lemma~\ref{lem:WiBoundedBelow}, there exists $\nu>0$ such that for all $\epsilon\in[0,\eta]$, $\nu I \leq \mathcal{W}_{\epsilon}(i,\tilde{X}_{\psi}(i,k+1,T,\epsilon)) = \mathcal{W}(i,\tilde{X}_{\psi}(i,k+1,T,\epsilon))$ for all $i\in\mathcal{N}$, $T\in\mathbb{N}_0$, $\psi\in\Psi$, $0\leq k\leq T$ where $X_{\psi}(i,T+1,T,\epsilon) = 0$ and
\begin{IEEEeqnarray}{rCl}
X_{\psi}(i,k,T,\epsilon) &=& \mathcal{S}_{\epsilon}(i,\tilde{X}_{\psi}(i,k+1,T,\epsilon)) \nonumber \\
&=& \mathcal{S}(i,\tilde{X}_{\psi}(i,k+1,T,\epsilon))+\epsilon I \label{eq:Seps} .
\end{IEEEeqnarray}
By Lemma~\ref{lem:RiccatiXkBoundedAboveUnif}, there exists $\rho>0$ such that for all $\epsilon\in[0,\eta]$, $0\leq X_{\psi}(i,k,T,\epsilon) \leq \rho I$ for all $i\in\mathcal{N}$, $T\in\mathbb{N}_0$, $\psi\in\Psi$, and $0\leq k \leq T+1$. That $X_{\psi}(i,k,T,\epsilon) \geq \epsilon I$ for $i\in\mathcal{N}$ and $0\leq k \leq T$ follows clearly from~\eqref{eq:Seps}.
\end{IEEEproof}

The following lemma allows comparison of the solutions of two Riccati difference equations in~\eqref{eq:smjlsRiccatiExistThmRecurDefAndFinalCond} with different values for $\epsilon$.
\begin{lem}[Lem. 2.6 of~\cite{lutzStilwell:leeDullerud2006Perf}]\label{lem:algIdents}
For $i\in\mathcal{N}$ and $X\in\mathbb{X}_i$, define
\begin{IEEEeqnarray}{c}
\mathcal{F}(i,X) = A(i) + B(i) \mathcal{W}^{-1}(i,X) \mathcal{R}(i,X). \label{eq:Fidef}
\end{IEEEeqnarray}
Let $Y\in\mathbb{X}_i$ and $\Delta=X-Y$. Then the following algebraic identities hold.
\begin{IEEEeqnarray}{rCl}
\IEEEeqnarraymulticol{3}{l}{
\mathcal{S}(i,X) - \mathcal{S}(i,Y)
} \nonumber \\ \quad
&=& \mathcal{F}^{\tpose}(i,Y)\Delta \mathcal{F}(i,Y) \nonumber \\
&& \quad +\, \mathcal{F}^{\tpose}(i,Y)\Delta B(i) \mathcal{W}^{-1}(i,X) B^{\tpose}(i) \Delta \mathcal{F}(i,Y) \IEEEeqnarraynumspace \label{eq:algIdentSymm}\\
&=& \mathcal{F}^{\tpose}(i,X) \Delta \mathcal{F}(i,Y) \label{eq:algIdentAsymm}
\end{IEEEeqnarray}
\end{lem}

Before proceeding, the following technical lemma is needed which is similar in nature to~\cite[Thm. 2.7(a)]{lutzStilwell:leeDullerud2006Perf}. The following lemma examines the random state transition matrix defined by
\begin{IEEEeqnarray}{rCl}
\phi(k,j,T)&=&\mathcal{F}(\theta(k-1),\tilde{X}_{\psi}(\theta(k-1),k,T,\epsilon)) \nonumber \\
&& \quad \times \cdots \times \mathcal{F}(\theta(j),\tilde{X}_{\psi}(\theta(j),j+1,T,\epsilon)) \IEEEeqnarraynumspace \label{eq:littlephidef}
\end{IEEEeqnarray}
when $k$ and $j$ are such that $0\leq j < k \leq T$, and $\phi(k,j,T)=I$ when $k=j$. Here, $\mathcal{F}(i,X)$ is defined as in~\eqref{eq:Fidef}, and $X_{\psi}$ is defined in~\eqref{eq:smjlsRiccatiExistThmRecurDefAndFinalCond} for a stable and contractive system $(\mathcal{G},\Pi,\Psi,p(0))$. Note that $\phi$ is only defined for $0\leq j \leq k \leq T$ and that dependence of $\phi$ on $\psi$ and $\epsilon$ is suppressed.
The state transition matrix in~\eqref{eq:littlephidef} arises from the recurrence $x(k+1) = \mathcal{F}(\theta(k),\tilde{X}_{\psi}(\theta(k),k+1,T,\epsilon))x(k)$, which is only defined for $0\leq k \leq T$.
\begin{lem}\label{lem:cEpsLambdaEpsExist}
Assume $p_i(k)>0$ for all $\psi\in\Psi$, $i\in\mathcal{N}$, and $k\in\mathbb{N}_0$. If $(\mathcal{G},\Pi,\Psi,p(0))$ is uniformly exponentially mean square stable and uniformly mean square strictly contractive then there exists $\eta>0$ such that for any $\epsilon\in(0,\eta)$ there exist $0\leq\lambda_\epsilon<1$ and $c_\epsilon>0$ such that 
\begin{IEEEeqnarray}{c}
\expec{\phi^{\tpose}(k,j,T)\phi(k,j,T) \mid \theta(j) = i} \leq c_\epsilon \lambda_\epsilon^{k-j} I \IEEEeqnarraynumspace \label{eq:FsystemEmssFinHoriz}
\end{IEEEeqnarray}
for all $i\in\mathcal{N}$, $T\in\mathbb{N}_0$, $\psi\in\Psi$, and all $0\leq j \leq k \leq T$ where $\phi$ is defined in~\eqref{eq:littlephidef}.
\end{lem}
\begin{IEEEproof}
Let $\eta>0$ be as in Theorem~\ref{thm:smjlsRiccatiExistThm} and fix $\epsilon\in(0,\eta)$. Fix $\epsilon_0\in(\epsilon,\eta)$ and let $\bar{\epsilon}=\epsilon_0 - \epsilon$. Define $Z_{\psi}(i,k,T,\bar{\epsilon}) := X_{\psi}(i,k,T,\epsilon_0) - X_{\psi}(i,k,T,\epsilon)$ and define $\tilde{Z}_{\psi}(i,k,T,\bar{\epsilon}) := \sum_{j=1}^N\pi_{ij}(\psi(k))Z_{\psi}(j,k,T,\bar{\epsilon})$ for $i\in\mathcal{N}$, $T\in\mathbb{N}_0$, $0\leq k\leq T+1$, $\psi\in\Psi$ where $X_{\psi}$ are solutions to~\eqref{eq:smjlsRiccatiExistThmRecurDefAndFinalCond}. Let $T\in\mathbb{N}_0$ and $\psi\in\Psi$ be arbitrary, and, for notational convenience, define $F(i,k) := \mathcal{F}(i,\tilde{X}_{\psi}(i,k+1,T,\epsilon))$ where $i\in\mathcal{N}$ and $0\leq k \leq T$. Then
\begin{IEEEeqnarray}{rCl}
\IEEEeqnarraymulticol{3}{l}{
Z_{\psi}(i,k,T,\bar{\epsilon})
}
\nonumber \\ \quad
&=& \mathcal{S}(i,\tilde{X}_{\psi}(i,k+1,T,\epsilon_0)) + \epsilon_0 I \nonumber \\
&& \quad -\, \mathcal{S}(i,\tilde{X}_{\psi}(i,k+1,T,\epsilon)) - \epsilon I \label{eq:eps2MinusEps1} \\
&\geq & F^{\tpose}(i,k) ( \tilde{X}_{\psi}(i,k+1,T,\epsilon_0) - \tilde{X}_{\psi}(i,k+1,T,\epsilon) ) \nonumber \\
&& \quad \times \, F(i,k) + \bar{\epsilon}I \IEEEeqnarraynumspace \label{eq:eps2MinusEps1DropPosSemDefTerms} \\
&=& F^{\tpose}(i,k)  \tilde{Z}_{\psi}(i,k+1,T,\bar{\epsilon})  F(i,k) + \bar{\epsilon}I \label{eq:eps2MinusEps1FinalRes}
\end{IEEEeqnarray}
where~\eqref{eq:eps2MinusEps1} follows from~\eqref{eq:smjlsRiccatiExistThmRecurDef}, and~\eqref{eq:eps2MinusEps1DropPosSemDefTerms} follows from~\eqref{eq:algIdentSymm}. Additionally, from~\eqref{eq:eps2MinusEps1FinalRes} and~\eqref{eq:smjlsRiccatiExistThmXikPosDefDecres}
\begin{IEEEeqnarray}{c}
\bar{\epsilon} I \leq Z_{\psi}(i,k,T,\bar{\epsilon}) \leq \bar{\rho} I \label{eq:eps2MinusEps1YkPosDefDecres}
\end{IEEEeqnarray}
for all $\psi\in\Psi$, $T\in\mathbb{N}_0$, $i\in\mathcal{N}$, and all $0\leq k \leq T$ where $\bar{\rho}=\rho - \epsilon$. Using stochastic Lyapunov function arguments as in the proof of~\cite{lutzStilwell:krtolica1994}[Thm. 2], inequalities~\eqref{eq:eps2MinusEps1FinalRes} and~\eqref{eq:eps2MinusEps1YkPosDefDecres} ensure that solutions to $x(k+1) = F(\theta(k),k)x(k)$ decay exponentially in mean square with $\lambda_\epsilon = 1 - \bar{\epsilon}/\bar{\rho}$ and $c_\epsilon = \bar{\rho}/\bar{\epsilon}$.
\end{IEEEproof}

We are now ready to provide an explicit construction for $Y_{\psi}$ in Lemma~\ref{lem:FinDependRicSolnTeaser}. The construction in the following lemma ensures that for each $k\in\mathbb{N}_0$, $Y_{\psi}(i,k)$ depends only on $i$ and $\psi_{M}(k)$.
\begin{lem}\label{lem:smjlsRiccatiFiniteDependence}
Assume $p_i(k)>0$ for all $\psi\in\Psi$, $i\in\mathcal{N}$, and $k\in\mathbb{N}_0$. If $(\mathcal{G},\Pi,\Psi,p(0))$ is uniformly exponentially mean square stable and uniformly mean square strictly contractive then there exist $\eta,\rho>0$ such that for all $\epsilon\in(0,\eta)$, there exist $M\in\mathbb{N}_0$ and $\nu>0$ such that $Y_{\psi}(i,k) := X_{\psi}(i,k,k+M,\epsilon)$ satisfies
\begin{IEEEeqnarray}{rCl}
\IEEEyesnumber \label{eq:smjlsFinHorBrl}
\epsilon I  \leq  Y_{\psi}(i,k) & \leq & \rho I \IEEEyessubnumber \label{eq:smjlsFinHorBrlA}
\\
\mathcal{B}(i,\tilde{Y}_{\psi}(i,k+1),Y_{\psi}(i,k))
& \leq & -\nu I
\IEEEyessubnumber \IEEEeqnarraynumspace \label{eq:smjlsFinHorBrlB}
\end{IEEEeqnarray}
for all $\psi\in\Psi$, $i\in\mathcal{N}$, and $k\in\mathbb{N}_0$ where $\tilde{Y}_{\psi}(i,k+1) = \sum_{j=1}^N \pi_{ij}(\psi(k+1)) Y_{\psi}(j,k+1)$ and $X_{\psi}$ is defined in~\eqref{eq:smjlsRiccatiExistThmRecurDefAndFinalCond}.
\end{lem}
\begin{IEEEproof}
Let $\eta,\rho$ be as in Theorem~\ref{thm:smjlsRiccatiExistThm} and choose $\epsilon\in(0,\eta)$ so that~\eqref{eq:smjlsFinHorBrlA} is verified automatically. Let $\lambda_\epsilon$ and $c_\epsilon$ be as in Lemma~\ref{lem:cEpsLambdaEpsExist}. Choose $M\in\mathbb{N}_0$ such that $c_\epsilon\lambda_\epsilon^{M+1}<\epsilon/\rho$. Then
\begin{IEEEeqnarray}{rCl}
\IEEEeqnarraymulticol{3}{l}{
\mathcal{S}(i,\tilde{Y}_{\psi}(i,k+1)) - Y_{\psi}(i,k) + \epsilon I
} \nonumber \\ \quad
&=&\mathcal{S}(i,\tilde{X}_{\psi}(i,k+1,k+M+1,\epsilon)) \nonumber \\
&&\quad -\, \mathcal{S}(i,\tilde{X}_{\psi}(i,k+1,k+M,\epsilon)) \nonumber \\ \quad
&=& \mathcal{F}^{\tpose}(i,\tilde{X}_{\psi}(i,k+1,k+M+1,\epsilon)) \nonumber \\
&&\quad \times \, \big(\tilde{X}_{\psi}(i,k+1,k+M+1,\epsilon) - \tilde{X}_{\psi}(i,k+1,k+M,\epsilon)\big) \nonumber \\
&& \quad \times \mathcal{F}(i,\tilde{X}_{\psi}(i,k+1,k+M,\epsilon)).\IEEEeqnarraynumspace \label{eq:firstSiDiff}
\end{IEEEeqnarray}
where~\eqref{eq:firstSiDiff} follows from~\eqref{eq:algIdentAsymm}. But the middle term in~\eqref{eq:firstSiDiff} can be written
\begin{IEEEeqnarray}{rCl}
\IEEEeqnarraymulticol{3}{l}{
\tilde{X}_{\psi}(i,k+1,k+M+1,\epsilon) - \tilde{X}_{\psi}(i,k+1,k+M,\epsilon) 
} \nonumber \\ 
&=& \sum_{j=1}^N \pi_{i j}(\psi(k+1)) 
\Big( \mathcal{S}(j,\tilde{X}_{\psi}(j,k+2,k+M+1,\epsilon)) \nonumber \\
&&\; -\, \mathcal{S}(j,\tilde{X}_{\psi}(j,k+2,k+M,\epsilon)) \Big)\IEEEeqnarraynumspace \label{eq:deltaTildeSi} \\
&=& \mathbf{E}\bigg[
\mathcal{F}^{\tpose}(\theta(k+1),\tilde{X}_{\psi}(\theta(k+1),k+2,k+M+1,\epsilon)) \nonumber \\
&& \; \times \Big( \tilde{X}_{\psi}(\theta(k+1),k+2,k+M+1,\epsilon) \nonumber \\
&& \; -\, \tilde{X}_{\psi}(\theta(k+1),k+2,k+M,\epsilon) \Big) \nonumber \\
&& \; \times \mathcal{F}(\theta(k+1),\tilde{X}_{\psi}(\theta(k+1),k+2,k+M,\epsilon)) \mid \theta(k)=i \bigg] \IEEEeqnarraynumspace \label{eq:deltaTildeExpecFi}
\end{IEEEeqnarray}
where~\eqref{eq:deltaTildeSi} follows from~\eqref{eq:smjlsRiccatiExistThmRecurDef}, and~\eqref{eq:deltaTildeExpecFi} results after applying~\eqref{eq:firstSiDiff} to~\eqref{eq:deltaTildeSi}. Proceeding in an iterative fashion, 
\begin{IEEEeqnarray}{rCl}
\IEEEeqnarraymulticol{3}{l}{
\mathcal{S}(i,\tilde{Y}_{\psi}(i,k+1)) - Y_{\psi}(i,k) + \epsilon I
} \nonumber \\ \quad
&=& \mathbf{E}\bigg[ \phi^{\tpose}(k+M+1,k,k+M+1) \nonumber \\
&& \quad \times \Big( \tilde{X}_{\psi}(\theta(k+M),k+M+1,k+M+1,\epsilon) \nonumber \\
&& \quad -\, \tilde{X}_{\psi}(\theta(k+M),k+M+1,k+M,\epsilon) \Big) \nonumber \\
&& \quad \times \phi(k+M+1,k,k+M) \mid \theta(k) = i \bigg]. \label{eq:SiDiffEqualExpec}
\end{IEEEeqnarray}
Note that the middle term in~\eqref{eq:SiDiffEqualExpec} satisfies
\begin{IEEEeqnarray}{c}
 \epsilon I \leq \tilde{X}_{\psi}(\theta(k+M),k+M+1,k+M+1,\epsilon) - 0 \leq \rho I \IEEEeqnarraynumspace \label{eq:XtildeBoundedByRho}
\end{IEEEeqnarray}
for all values of $\theta(k+M)\in\mathcal{N}$.
Let $y\in\mathbb{R}^n$ be arbitrary, and for convenience define $\phi_1 = \phi(k+M+1,k,k+M+1)$, $\phi_2 = \phi(k+M+1,k,k+M)$, and $X= \tilde{X}_{\psi}(\theta(k+M),k+M+1,k+M+1,\epsilon)$. Then
\begin{IEEEeqnarray}{rCl}
\IEEEeqnarraymulticol{3}{l}{
y^{\tpose} \left( \mathcal{S}(i,\tilde{Y}_{\psi}(i,k+1)) - Y_{\psi}(i,k) + \epsilon I \right) y
} \nonumber \\ \quad
&=& y^{\tpose} \expec{  \phi_{1}^{\tpose} X \phi_2  \mid \theta(k) = i } y \nonumber \\
& \leq & \sqrt{ \expec{ v_{1}^{\tpose} X v_{1} \mid \theta(k)=i } \expec{ v_{2}^{\tpose} X v_{2} \mid \theta(k)=i } } \label{eq:applyGeneralCSineq} \\
& \leq & \sqrt{ \rho^{2} \expec{ v_{1}^{\tpose} v_{1} \mid \theta(k)=i } \expec{ v_{2}^{\tpose} v_{2} \mid \theta(k)=i } } \label{eq:XupperBdRho} \\
&\leq & \rho \sqrt{ \left( c_{\epsilon} \lambda_{\epsilon}^{M+1} y^{\tpose}  y \right) \left( c_{\epsilon} \lambda_{\epsilon}^{M+1} y^{\tpose}  y \right) } \label{eq:applyPhiDecayLemma} \\
 &<& \epsilon y^{\tpose} y \label{eq:choiceOfMyieldsEpsilon}
\end{IEEEeqnarray}
where $v_{1} = \phi_{1} y$ and $v_2 = \phi_{2} y$; \eqref{eq:applyGeneralCSineq} follows from $X>0$ in~\eqref{eq:XtildeBoundedByRho} and the Cauchy-Schwarz inequality for random variables; \eqref{eq:XupperBdRho} follows from~\eqref{eq:XtildeBoundedByRho}; \eqref{eq:applyPhiDecayLemma} follows from Lemma~\ref{lem:cEpsLambdaEpsExist}; and, \eqref{eq:choiceOfMyieldsEpsilon} follows by choice of $M$. 
Let $\epsilon_0\in(0,\epsilon)$ be such that
\begin{IEEEeqnarray}{c}
\mathcal{S}(i,\tilde{Y}_{\psi}(i,k+1)) - Y_{\psi}(i,k) + \epsilon I \leq \epsilon_0 I < \epsilon I. \label{eq:SiDiffLeqEps0}\IEEEeqnarraynumspace
\end{IEEEeqnarray}
By~\eqref{eq:SiDiffLeqEps0}, $\mathcal{S}(i,\tilde{Y}_{\psi}(i,k+1)) - Y_{\psi}(i,k) \leq -\nu I$ where $\nu=\epsilon-\epsilon_0$. Application of the Schur complement yields~\eqref{eq:smjlsFinHorBrlB}.
\end{IEEEproof}

Lemma~\ref{lem:smjlsRiccatiFiniteDependence} uses techniques similar to those found in~\cite[Thm. 2.7(b)]{lutzStilwell:leeDullerud2006Perf} where it is shown that the time-varying version of the KYP inequality associated with a uniformly stable and contractive linear time-varying system admits a solution with finite memory of past parameters.

\begin{rem}\label{rmk:smjlsRiccatiFiniteDependenceRmk}
The construction in Lemma~\ref{lem:smjlsRiccatiFiniteDependence} ensures that $Y_{\psi}(i,k)$ may be computed with knowledge of only $i$ and $\psi_{M}(k)$. Indeed, if $t\neq k$ but $\psi_{M}(k)=\psi_{M}(t)$ then $Y_{\psi}(i,k)=Y_{\psi}(i,t)$. This claim can be easily established using the recursive relation~\eqref{eq:smjlsRiccatiExistThmRecurDef} and base case $X_{\psi}(i,k+M+1,k+M,\epsilon) = X_{\psi}(i,t+M+1,t+M,\epsilon) = 0$.
\end{rem}

The following theorem, inspired by~\cite[Thm. 3.3]{lutzStilwell:leeDullerud2006Perf}, provides a necessary and sufficient condition, expressed as a set of finite-dimensional LMIs, for uniform exponential mean square stability and uniform mean square strict contractiveness of a switched Markov jump linear system.
\begin{thm}\label{thm:smjlsUMSSC}
Assume $p_i(k)>0$ for all $\psi\in\Psi$, $i\in\mathcal{N}$, and $k\in\mathbb{N}_0$. The switched Markov jump linear system $(\mathcal{G},\Pi,\Psi,p(0))$ is uniformly exponentially mean square stable and uniformly mean square strictly contractive if and only if there exist $M\in\mathbb{N}_0$ and a function $X:\mathcal{N}\times\Psi_{M}\to\mathbb{S}_{n}^{+}$ such that for any $(r_1,\dots,r_{M+1})\in\Psi_{M+1}$ and $i\in\mathcal{N}$
\begin{IEEEeqnarray}{c}
\mathcal{B}\left( i,\sum_{j=1}^N \pi_{ij}(r_1) X(j,r_2,\dots, r_{M+1}),X(i,r_1,\dots, r_{M}) \right) < 0 \IEEEeqnarraynumspace \label{eq:smjUMSC}
\end{IEEEeqnarray}
\end{thm}
\begin{IEEEproof}
Suppose there exist $M$ and $X$ such that~\eqref{eq:smjUMSC} holds. Note that the upper left block of~\eqref{eq:smjUMSC} implies~\eqref{eq:smjUEMSS} so uniform exponential mean square stability of $(\mathcal{G},\Pi,\Psi,p(0))$ follows from Theorem~\ref{thm:smjUEMSS}. Since $\mathcal{N}\times\Psi_{M+1}\subset \mathcal{N}\times\mathcal{J}^{M+1}$ is a finite set, inequality~\eqref{eq:smjUMSC} holds uniformly, and we can find $0<\nu<1$ such that $\mathcal{B}\left( i,\sum_{j=1}^N \pi_{ij}(r_1) X(j,r_2,\dots, r_{M+1}),X(i,r_1,\dots, r_{M}) \right)
\leq  -\nu I$ for any $(i,r_1,\dots,r_{M+1})\in\mathcal{N}\times\Psi_{M+1}$. Let $\psi\in\Psi$ be arbitrary. Define $Y_{\psi}(i,k):=X(i,\psi_{M}(k))$. Using~\eqref{eq:MjlsTihBrlBDiffForm} and \eqref{eq:l2NormMatrixIdent} to rewrite $\mathcal{B}$, it follows that
\begin{IEEEeqnarray}{rCl}
\IEEEeqnarraymulticol{3}{l}{
\mathbf{E}\big[ \, \lVert z(k) \rVert^2 + x^{\tpose}(k+1) Y_{\psi}(\theta(k+1),k+1) x(k+1)
}
\nonumber \\
&& \quad -\, x^{\tpose}(k) Y_{\psi}(\theta(k),k) x(k) \, \big]  
\leq (1-\nu)\mathbf{E}[ \, \lVert w(k)\rVert^2 \, ].\IEEEeqnarraynumspace \label{eq:zkwkIneq}
\end{IEEEeqnarray}
Inequality~\eqref{eq:zkwkIneq}, positive definiteness of $Y_{\psi}(i,k)$, and $x(0)=0$ imply $\sum_{k=0}^{l} \expec{\norm{z(k)}^2} \leq (1-\nu)\sum_{k=0}^{l} \expec{\norm{w(k)}^2}$ for all $l\in\mathbb{N}_0$. Since $\psi\in\Psi$ was arbitrary, Definition~\ref{defn:umssc} is satisfied with $\gamma=\sqrt{1-\nu}$ so $(\mathcal{G},\Pi,\Psi,p(0))$ is uniformly mean square strictly contractive.

Conversely, assume that $(\mathcal{G},\Pi,\Psi,p(0))$ is uniformly exponentially mean square stable and uniformly mean square strictly contractive. Let $\eta,\rho$ be as in Lemma~\ref{lem:smjlsRiccatiFiniteDependence}, fix $\epsilon\in(0,\eta)$, and let $M$ and $\nu$ be defined as in Lemma~\ref{lem:smjlsRiccatiFiniteDependence}. Let $(i,r_1,\dots,r_{M+1})\in\mathcal{N}\times\Psi_{M+1}$ be arbitrary. By definition of $\Psi_{M+1}$, there exist $\psi\in\Psi$ and $t\in\mathbb{N}_0$ such that $\psi_{M+1}(t) = (r_1,\dots,r_{M+1})$. Construct $Y_{\psi}$ as in Lemma~\ref{lem:smjlsRiccatiFiniteDependence} and recall from Remark~\ref{rmk:smjlsRiccatiFiniteDependenceRmk} that $Y_{\psi}(i,t)$ depends only on $(i,\psi_{M}(t))$. Thus, define $X(i,r_1,\dots,r_M):= Y_{\psi}(i,t)$ and define $X(i,r_2,\dots,r_{M+1}):=Y_{\psi}(i,t+1)$. One recovers every inequality in~\eqref{eq:smjUMSC} from~\eqref{eq:smjlsFinHorBrl}.
\end{IEEEproof}

\begin{rem}
Theorem~\ref{thm:smjlsUMSSC} provides a practical approach for investigating the contractiveness of a \emph{single} time-inhomogeneous Markov jump linear system with \emph{known} transition probability matrices that vary in a finite set (let $\Psi$ be the set containing a single sequence).
\end{rem}

\begin{rem}
Consider the case when $J=1$ and $\Psi = \{ (1,1,\dots) \}$. The switched Markov jump linear system $(\mathcal{G},\Pi,\Psi,p(0))$ reduces to a single time-homogeneous Markov jump linear system $(\mathcal{G},\Pi\circ \psi_1,p(0))$ where $\psi_1\equiv 1$. For any $M$, the set $\Psi_{M}$ contains only a single element $(1,\dots,1)$, and the set $\mathcal{N}\times\Psi_{M}$ contains only $N$ elements. For $i\in\mathcal{N}$, define $Z(i):= X(i,1,\dots,1)$ where $X$ is as in Theorem~\ref{thm:smjlsUMSSC}. Then~\eqref{eq:smjUMSC} reduces to $\mathcal{B}( i,\tilde{Z}(i),Z(i) ) <0$ where $\tilde{Z}(i)=\sum_{j=1}^N \pi_{ij}(1) Z(j)$, which is the same inequality found in the well-known bounded real lemma for time-homogeneous Markov jump linear systems~\cite[Thm. 2]{lutzStilwell:seiler2003}. Theorem~\ref{thm:smjlsUMSSC}, however, does not require $(\mathcal{G},\Pi\circ\psi_1,p(0))$ to be weakly controllable as in~\cite{lutzStilwell:seiler2003}. Thus, the weak controllability hypothesis of~\cite[Thm. 2]{lutzStilwell:seiler2003} can be replaced by the weaker (see Proposition~\ref{prop:weakControlStrongerCondit}) hypothesis that $p_i(k)>0$ for all $i\in\mathcal{N}$ and all $k\in\mathbb{N}_0$. Recalling Proposition~\ref{prop:probHypToInitDistAndPkHyp}, this hypothesis is equivalent to $p_i(0)>0$ for all $i\in\mathcal{N}$, and each column of $\Pi(1)$ is nonzero.
\end{rem}

\begin{prop}\label{prop:weakControlStrongerCondit}
Let $\Pi(1)$ be a stochastic matrix and $\psi_1\equiv 1$. If the time-homogeneous Markov jump linear system $(\mathcal{G},\Pi\circ\psi_1,p(0))$ is weakly controllable and $p_i(0)>0$ for all $i\in\mathcal{N}$, then $p_i(k)>0$ for all $i\in\mathcal{N}$ and $k\in\mathbb{N}_0$.
\end{prop}
\begin{IEEEproof}
The contrapositive is proved. Suppose the conclusion of the conditional statement is false. Then by Proposition~\ref{prop:probHypToInitDistAndPkHyp}, $\Pi(1)$ has a zero column and/or $p_i(0)=0$ for some $i\in\mathcal{N}$. If the $j$-th column of $\Pi(1)$ is zero, then $\pi_{ij}(1)=0$ for all $i\in\mathcal{N}$ and $\prob{x(k) = x_f, \theta(k) = j}\leq\prob{\theta(k) = j}=\sum_{i=1}^{N} \pi_{ij}(1)\prob{\theta(k-1)=i}=0$ for all $k\geq 1$. Thus, the final state $(x_f,j)$ has zero probability for all $k\geq 1$ and any input $w_c\in\ell^2_e$ so $(\mathcal{G},\Pi\circ\psi_1,p(0))$ is not weakly controllable.
\end{IEEEproof}

\section{Examples}
\begin{example}\label{example:perfExample}
Consider the switched Markov jump linear system $(\mathcal{G},\Pi,\Psi,p(0))$ where
\begin{IEEEeqnarray*}{rCl.rCl}
A(1) &=&
\begin{bmatrix}
0.08 & 0.15 & 0.30 \\ 
0.20 & 0.60 & 0.10 \\
0.50 & 0.20 & 0.40 
\end{bmatrix},
&
B(1) &=&
\begin{bmatrix}
0.10 & 0.70 \\
0.50 & 0.80 \\
0.20 & 0.40
\end{bmatrix},
\\
C(1) &=& 
\begin{bmatrix}
0.18 & 0.03 & 0.01 \\
0.01 & 0.07 & 0.06 \\
0.02 & 0.03 & 0.15 
\end{bmatrix},
&
D(1) &=& 
\begin{bmatrix}
0.01 & 0 \\
0.08 & 0.05 \\
0 & 0.01 
\end{bmatrix},
\\
A(2) &=&
\begin{bmatrix}
-0.06 & 0.40 & 0.70 \\
0.35 & -0.07 & 0.10 \\
0.23 & -0.04 & 0.51
\end{bmatrix},
&
B(2) &=&
\begin{bmatrix}
0.41 & -0.75 \\
0.90 & 0.47 \\
0.54 & 0.28
\end{bmatrix},
\\
C(2) &=& 
\begin{bmatrix}
0.03 & -0.02 & 0.03 \\
0.07 & 0.09 & 0.10 \\
0.07 & 0.02 & 0.08
\end{bmatrix},
&
D(2) &=& 
\begin{bmatrix}
0 & 0.03 \\
0.01 & -0.11 \\ 
0 & 0.05
\end{bmatrix}, 
\\
\Pi(1) & = &
\begin{bmatrix}
0.46 & 0.54 \\
0.40 & 0.60 
\end{bmatrix},
&
\Pi(2) & = &
\begin{bmatrix}
0.01 & 0.99 \\
0.05 & 0.95
\end{bmatrix}
\end{IEEEeqnarray*}
$p(0) = [ 0.5 \, 0.5 ]$, and $\Psi$ is the set of all sequences in $\mathcal{J}=\{1,2\}$. Note that the linear time-invariant system with fixed matrices $(A(1),B(1),C(1),D(1))$ is exponentially stable but not contractive. On the other hand, the linear time-invariant system with fixed matrices $(A(2),\allowbreak B(2),\allowbreak C(2),\allowbreak D(2))$ is exponentially stable and contractive. The LMIs in Theorem~\ref{thm:smjlsUMSSC} are not feasible with $M=0$, but are feasible with $M=1$. Thus, the switched Markov jump linear system $(\mathcal{G},\Pi,\Psi,p(0))$ is uniformly exponentially mean square stable and uniformly mean square strictly contractive. For this work, YALMIP \cite{lutzStilwell:yalmip} was used with SeDuMi \cite{lutzStilwell:sedumi} to solve the convex feasibility problem.
\end{example}

\begin{example}\label{ex:switchPathConstrEx}
Now consider the switched Markov jump linear system $(\mathcal{G},\Pi,\Psi,p(0))$ with $A(1)$ and $p(0)$ defined in Example~\ref{example:perfExample} and
\begin{IEEEeqnarray*}{c}
A(2) =
\begin{bmatrix}
0.30 & 0.60 & 0.50 \\
0.30 & 0.70 & 0.20 \\
0.90 & 0.70 & 0.10
\end{bmatrix},
\nonumber \\
\Pi(1) = 
\begin{bmatrix}
0.90 & 0.10 \\
0.80 & 0.20 
\end{bmatrix},
\quad
\Pi(2) = 
\begin{bmatrix}
0.60 & 0.40 \\
0.75 & 0.25
\end{bmatrix} .
\end{IEEEeqnarray*}
Let $\Psi$ be the set of sequences that could arise from traversing the directed graph in Fig.~\ref{fig:directedGraphForSwitchPathConstrEx}.
\begin{figure}[t]
\centering
\begin{tikzpicture}[->,>=stealth',shorten >=1pt,auto,node distance=1.5cm,
  thick,main node/.style={circle,draw}]

  \node[main node] (1) {$1$};
  \node[main node] (2) [right of=1] {$2$};

  \path[]
    (1) edge [bend left] (2)
        edge [loop left] (1)
    (2) edge [bend left] (1);
\end{tikzpicture}
\caption{Directed graph that determines $\Psi$ in Example~\ref{ex:switchPathConstrEx}.}
\label{fig:directedGraphForSwitchPathConstrEx}
\end{figure}
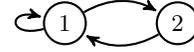
Note that matrix $A(2)$ is not Schur and that $\Pi(2)$ places a larger conditional probability on $\theta(k)=2$ than $\Pi(1)$. Solving the convex feasibility problem posed in Remark~\ref{rem:stabReducesToThWhenSingleton}, the time-homogeneous Markov jump linear system $(\mathcal{G},\Pi\circ\psi_2,p(0))$ where $\psi_2\equiv 2$ is \emph{not} exponentially mean square stable, while the time-homogeneous Markov jump linear system $(\mathcal{G},\Pi\circ\psi_1,p(0))$ where $\psi_1\equiv 1$ is exponentially mean square stable. Note from Fig.~\ref{fig:directedGraphForSwitchPathConstrEx} that $\psi_1\in\Psi$ while $\psi_2\not\in\Psi$. The LMIs in Theorem~\ref{thm:smjUEMSS} are feasible with $M=1$ so $(\mathcal{G},\Pi,\Psi,p(0))$ is uniformly exponentially mean square stable. Thus, uniform stability may sometimes be gained by imposing constraints on the switching set~$\Psi$.
\end{example}

\section{Conclusion}
This paper considers uniform exponential mean square stability and uniform mean square strict contractiveness of a switched Markov jump linear system. The switched Markov jump linear system abstraction is useful when transition probability matrices of the Markov chain vary with time in an a priori unknown manner. The mean square concepts examined are appropriate when at least one subsystem $(A(i),B(i),C(i),D(i))$ is not stable or not contractive when considered as a linear time-invariant system. Necessary and sufficient conditions for a switched Markov jump linear system to be uniformly exponentially mean square stable and uniformly mean square strictly contractive were developed. The conditions are convex and can be directly applied in practice.

\ifCLASSOPTIONcaptionsoff
  \newpage
\fi

\bibliographystyle{IEEEtran}

\bibliography{IEEEabrv,lutzStilwellBib}

\begin{thebibliography}{10}
\providecommand{\url}[1]{#1}
\csname url@samestyle\endcsname
\providecommand{\newblock}{\relax}
\providecommand{\bibinfo}[2]{#2}
\providecommand{\BIBentrySTDinterwordspacing}{\spaceskip=0pt\relax}
\providecommand{\BIBentryALTinterwordstretchfactor}{4}
\providecommand{\BIBentryALTinterwordspacing}{\spaceskip=\fontdimen2\font plus
\BIBentryALTinterwordstretchfactor\fontdimen3\font minus
  \fontdimen4\font\relax}
\providecommand{\BIBforeignlanguage}[2]{{%
\expandafter\ifx\csname l@#1\endcsname\relax
\typeout{** WARNING: IEEEtran.bst: No hyphenation pattern has been}%
\typeout{** loaded for the language `#1'. Using the pattern for}%
\typeout{** the default language instead.}%
\else
\language=\csname l@#1\endcsname
\fi
#2}}
\providecommand{\BIBdecl}{\relax}
\BIBdecl

\bibitem{lutzStilwell:doVal1999}
J.~B. do~Val and T.~Ba{\c{s}}ar, ``Receding horizon control of jump linear
  systems and a macroeconomic policy problem,'' \emph{J. Econ. Dyn. Control},
  vol.~23, no.~8, pp. 1099--1131, 1999.

\bibitem{lutzStilwell:chizeck1982diss}
H.~J. Chizeck, ``Fault tolerant optimal control,'' {Sc.D.} dissertation,
  Massachusetts Inst. Technol., Cambridge, MA, 1982.

\bibitem{lutzStilwell:lutz2013}
C.~C. Lutz and D.~J. Stilwell, ``Energy-aware control: $\ell_2$ gain for
  closed-loop systems implemented with stochastic schedulers,'' in \emph{Proc.
  Amer. Control Conf.}\hskip 1em plus 0.5em minus 0.4em\relax Washington, DC,
  USA: IEEE, 2013, pp. 5313--5319.

\bibitem{lutzStilwell:hespanha2007}
J.~P. Hespanha, P.~Naghshtabrizi, and Y.~Xu, ``A survey of recent results in
  networked control systems,'' \emph{Proc. {IEEE}}, vol.~95, no.~1, pp.
  138--162, 2007.

\bibitem{lutzStilwell:seiler2005}
P.~Seiler and R.~Sengupta, ``An ${H}_\infty$ approach to networked control,''
  \emph{{IEEE} Trans. Autom. Control}, vol.~50, no.~3, pp. 356--364, 2005.

\bibitem{lutzStilwell:xiao2000}
L.~Xiao, A.~Hassibi, and J.~P. How, ``Control with random communication delays
  via a discrete-time jump system approach,'' in \emph{Proc. Amer. Control
  Conf.}\hskip 1em plus 0.5em minus 0.4em\relax Chicago, IL, USA: IEEE, 2000,
  pp. 2199--2204.

\bibitem{lutzStilwell:ploplys2004}
N.~J. Ploplys, P.~A. Kawka, and A.~G. Alleyne, ``Closed-loop control over
  wireless networks,'' \emph{{IEEE} Control Syst. Mag.}, vol.~24, no.~3, pp.
  58--71, 2004.

\bibitem{lutzStilwell:ji1991}
Y.~Ji, H.~J. Chizeck, X.~Feng, and K.~A. Loparo, ``Stability and control of
  discrete-time jump linear systems,'' \emph{Control Theory Adv. Technol.},
  vol.~7, no.~2, pp. 247--270, Jun. 1991.

\bibitem{lutzStilwell:costa1993}
O.~L. Costa and M.~D. Fragoso, ``Stability results for discrete-time linear
  systems with {M}arkovian jumping parameters,'' \emph{J. Math. Anal. Appl.},
  vol. 179, no.~1, pp. 154--178, 1993.

\bibitem{lutzStilwell:ji1990a}
Y.~Ji and H.~J. Chizeck, ``Jump linear quadratic {G}aussian control:
  steady-state solution and testable conditions,'' \emph{Control Theory Adv.
  Technol.}, vol.~6, no.~3, pp. 289--319, Sep. 1990.

\bibitem{lutzStilwell:seiler2003}
P.~Seiler and R.~Sengupta, ``A bounded real lemma for jump systems,''
  \emph{{IEEE} Trans. Autom. Control}, vol.~48, no.~9, pp. 1651--1654, Sep.
  2003.

\bibitem{lutzStilwell:leeDullerud2006Stab}
J.-W. Lee and G.~E. Dullerud, ``Uniform stabilization of discrete-time switched
  and {M}arkovian jump linear systems,'' \emph{Automatica}, vol.~42, no.~2, pp.
  205--218, 2006.

\bibitem{lutzStilwell:leeDullerud2006Perf}
------, ``Optimal disturbance attenuation for discrete-time switched and
  {M}arkovian jump linear systems,'' \emph{SIAM J. Control Optim.}, vol.~45,
  no.~4, pp. 1329--1358, 2006.

\bibitem{lutzStilwell:krtolica1994}
R.~Krtolica, {\"U}.~{\"O}zg{\"u}ner, H.~Chan, H.~G{\"o}ktas, J.~Winkelman, and
  M.~Liubakka, ``Stability of linear feedback systems with random communication
  delays,'' \emph{Int. J. Control}, vol.~59, no.~4, pp. 925--953, 1994.

\bibitem{lutzStilwell:aberkane2013}
S.~Aberkane, ``Bounded real lemma for nonhomogeneous {M}arkovian jump linear
  systems,'' \emph{{IEEE} Trans. Autom. Control}, vol.~58, no.~3, pp. 797--801,
  Mar. 2013.

\bibitem{lutzStilwell:fang2002}
Y.~Fang and K.~A. Loparo, ``Stochastic stability of jump linear systems,''
  \emph{{IEEE} Trans. Autom. Control}, vol.~47, no.~7, pp. 1204--1208, 2002.

\bibitem{lutzStilwell:bolzern2010}
P.~Bolzern, P.~Colaneri, and G.~De~Nicolao, ``{M}arkov jump linear systems with
  switching transition rates: mean square stability with dwell-time,''
  \emph{Automatica}, vol.~46, no.~6, pp. 1081--1088, 2010.

\bibitem{lutzStilwell:garcia2008book}
A.~Leon-Garcia, \emph{Probability, Statistics, and Random Processes for
  Electrical Engineering}, 3rd~ed.\hskip 1em plus 0.5em minus 0.4em\relax
  Pearson Education, 2008.

\bibitem{lutzStilwell:hrbacek1999book}
K.~Hrbacek and T.~Jech, \emph{Introduction to Set Theory}, 3rd~ed.\hskip 1em
  plus 0.5em minus 0.4em\relax Marcel Dekker, 1999.

\bibitem{lutzStilwell:rugh1996book}
W.~J. Rugh, \emph{Linear system theory}.\hskip 1em plus 0.5em minus 0.4em\relax
  Prentice-Hall, 1996.

\bibitem{lutzStilwell:dullerud1999}
G.~E. Dullerud and S.~Lall, ``A new approach for analysis and synthesis of
  time-varying systems,'' \emph{{IEEE} Trans. Autom. Control}, vol.~44, no.~8,
  pp. 1486--1497, Aug. 1999.

\bibitem{lutzStilwell:willems1972i}
J.~C. Willems, ``Dissipative dynamical systems part {I}: General theory,''
  \emph{Arch. Rational Mech. Anal.}, vol.~45, no.~5, pp. 321--351, 1972.

\bibitem{lutzStilwell:liberzon2003book}
D.~Liberzon, \emph{Switching in systems and control}.\hskip 1em plus 0.5em
  minus 0.4em\relax Birkh{\"a}user Boston, 2003.

\bibitem{lutzStilwell:limebeer1989}
D.~Limebeer, M.~Green, and D.~Walker, ``Discrete time ${H}^\infty$ control,''
  in \emph{Proc. 28th IEEE Conf. Decision and Control}.\hskip 1em plus 0.5em
  minus 0.4em\relax Tampa, FL, USA: IEEE, 1989, pp. 392--396.

\bibitem{lutzStilwell:greenLimebeer1995}
M.~Green and D.~J. Limebeer, \emph{Linear robust control}.\hskip 1em plus 0.5em
  minus 0.4em\relax Prentice-Hall, 1995.

\bibitem{lutzStilwell:yalmip}
\BIBentryALTinterwordspacing
J.~L\"{o}fberg, ``{YALMIP} : A toolbox for modeling and optimization in
  {MATLAB},'' in \emph{Proc. {IEEE} Int. Symp. Computer-Aided Control Syst.
  Des.}\hskip 1em plus 0.5em minus 0.4em\relax Taipei, Taiwan: IEEE, 2004, pp.
  284--289. [Online]. Available: \url{http://users.isy.liu.se/johanl/yalmip}
\BIBentrySTDinterwordspacing

\bibitem{lutzStilwell:sedumi}
J.~F. Sturm, ``Using {SeDuMi} 1.02, a {MATLAB} toolbox for optimization over
  symmetric cones,'' \emph{Optimization Meth. \& Soft.}, vol.~11, no. 1-4, pp.
  625--653, 1999.

\end{thebibliography}

\end{document}